\documentclass[final,3p]{elsarticle}

\usepackage{amsmath,amssymb}
\usepackage{hyperref}

\journal{Nuclear Physics B}
\begin{document}

\begin{frontmatter}

\title{Neutrino Mass Hierarchy and Axion--Lepton Couplings in a Minimal Flavored $\nu$KSVZ Model}

\author[hust]{Zhen-Yu Lei}
\author[hust]{Jian-Wei Cui\footnote{jwcui@hust.edu.cn}}
\affiliation[hust]{
			organization={School of Physics, Huazhong University of Science and Technology},
            city={Wuhan},
            country={China}
            }

\begin{abstract}

By treating the PQ symmetry as a flavor symmetry, we construct a minimal flavored axion model by combining the KSVZ axion framework with the type-I seesaw mechanism. This minimal setup introduces flavor-dependent PQ charge assignments and two additional SM singlet scalar fields, leading to a specific zero texture in the inverse Majorana neutrino mass matrix. This texture, together with the observed smallness of $\theta_{13}$, predicts an approximate inverted GST-like relation, implying an inverted neutrino mass ordering with $17\lesssim m_1/m_3\lesssim45$. Assuming no strong hierarchy among the relevant Majorana Yukawa couplings, this ordering is naturally determined by the ratio of the vacuum expectation values of the two SM-singlet scalar fields. 
This predicted inverted structure is the most readily testable feature of our model, given that current experiments favor the normal ordering. Nevertheless, since the inverted one has not been ruled out, a viable parameter space still remains.
We also perform a comprehensive study of the axion--lepton interactions in this framework. The induced axion--neutrino couplings satisfy current astrophysical constraints while potentially affecting neutrino oscillations under suitable conditions. Furthermore, we derive analytical expressions for the axion couplings to charged leptons by evaluating the one-loop radiative corrections in a largely model independent framework and assess the corresponding experimental constraints.

\end{abstract}


\end{frontmatter}

\tableofcontents

\section{Introduction}
\label{sec:intro}
An interesting and natural approach to solving the strong CP problem is to introduce a global chiral $U(1)$ symmetry known as the Peccei--Quinn (PQ) symmetry \cite{PQ1,PQ2}. The spontaneous breaking of this PQ symmetry gives rise to a pseudo Goldstone boson, the axion \cite{PhysRevLett.40.223,PhysRevLett.40.279,Kim-KSVZ,SHIFMAN1980493,DINE1981199}, which rolls to the minimum of its effective potential and settles the strong CP phase extremely close to 0. The axion can also serve as a dark matter candidate \cite{Preskill1983,Abbott1983,Dine1983,TURNER199067,kim2010review}.
	
On the other hand, the well-known seesaw mechanism provides a natural explanation for the smallness of neutrino masses by introducing heavy right-handed Majorana neutrinos \cite{Minkowski1977,Gell-Mann1979,Yanagida1979,PhysRevLett.44.912,PhysRevD.22.2227,Glashow1980}.
An intriguing coincidence is that the characteristic scales of the seesaw mechanism and invisible axion models lie in a similar range, approximately $M^{(R)}\sim f_a\sim 10^9\text{--}10^{12} $ GeV \cite{DILUZIO20201}. This coincidence naturally suggests the possibility of a connection between the two frameworks.
Previous works~\cite{Shin:1987xc,Dias2014,Ballesteros2017,Ballesteros2017_2,Giannotti2017,Pilaftsis_1994,Cely2017} and~\cite{Langacker1986,Volkas1988,PhysRevD.93.035001,Matlis2024,RVolkas2023} have constructed the so-called $\nu$--Kim--Shifman--Vainshtein--Zakharov ($\nu$KSVZ) and $\nu$--Dine--Fischler--Srednicki--Zhitnitsky ($\nu$DFZS) models respectively, in which the PQ charges are assigned in a flavor-universal way, naturally connecting the axion with the seesaw mechanism.
These models could solve the strong CP problem and simultaneously generate light neutrino masses.
	
When the PQ symmetry is treated as a flavor symmetry, it can also induce specific structures in fermion mass matrices \cite{Berezhiani1991,Davidson1984647,Davidson:1981zd, Davidson:1983fe,Bjrkeroth2019,Cox2024}. Consequently, flavored axion--neutrino models may lead to distinctive textures in the neutrino mass matrix. 
In this context, Ref. \cite{Bj_rkeroth_2020} investigated such textures arising from a ``full horizontality'' \cite{Davidson1984647} U(1) symmetry which can be interpreted as a (DFSZ-like) PQ symmetry.
Along these lines, recent investigations into flavored $\nu$DFSZ models \cite{Rocha2025, Giraldo2026, Babu2026} have further explored the nexus between neutrino phenomenology, axion dynamics, and the origin of flavor structures.

In this work, we combine these ideas within the KSVZ framework by constructing a minimal flavored $\nu$KSVZ model.
In this model, a zero texture appears in the inverse neutrino mass matrix, $(M_\nu^{-1})_{11} = 0$. Combined with the current neutrino oscillation data, this texture leads to an approximate inverted Gatto--Sartori--Tonin (GST) -like relation \cite{Gatto1968,Fritzsch1977,Roy2020} $s_{13}^2\sim m_3/m_1$, which implies an Inverted Hierarchy (IH) of neutrino masses with $17 \lesssim m_1/m_3 \lesssim 45$. Although the inverted neutrino mass ordering is currently disfavored by current oscillation data, it is not excluded. Consequently, it constitutes the most experimentally testable prediction of the present model.
Furthermore, we show that, provided there is no strong hierarchy among the relevant Majorana Yukawa couplings, $\sin\theta_{13}$ is approximately determined by the ratio of the vacuum expectation values (VEVs) of the two Standard Model (SM) singlet scalars. Since there is no theoretical requirement that these VEVs be of comparable magnitude, the smallness of $\theta_{13}$ emerges naturally.

Due to the non-universal PQ charge assignments, the axion--neutrino interactions in this model naturally acquire off-diagonal components in the mass basis, which are subject to the most stringent astrophysical constraints.  
Beyond these astrophysical constraints, such couplings may also influence neutrino oscillations in certain scenarios~\cite{huang2018neutrinophilic, losada2023parametric}.
In addition to their role in the neutrino sector, axion couplings to charged leptons are also important probes of axion physics \cite{Bertolami2014,Giannotti2017,DILUZIO20201,Bollig2020,Ghosh2020,Bauer_2022}. In this work, we derive analytical expressions for these couplings up to the one-loop level under minimal assumptions, making the results applicable to a broad class of $\nu$KSVZ models which satisfy two essential assumptions. 
Although the couplings are more involved than in the flavor-universal models \cite{Shin:1987xc,Cely2017,Heeck2019}, we show that the diagonal CP-violating terms \cite{Raffelt2012,Hare2020} still vanish at leading order in the seesaw expansion. The resulting couplings therefore remain consistent with existing experimental constraints.

The remainder of this paper is organized as follows. In Sec.~\ref{sec:model}, we construct a minimal flavored $\nu$KSVZ model and derive the resulting texture of the neutrino mass matrix.
In Sec.~\ref{sec:hierarchy}, we analyze the phenomenological implications of this texture, showing that the model predicts an inverted neutrino mass hierarchy together with a hierarchy between the VEVs of the two SM singlet scalars. The effective Majorana mass $\langle m_{\beta\beta}\rangle$ is also derived.
In Sec.~\ref{sec:couplings}, we investigate the axion--neutrino interactions and their phenomenological implications, and subsequently derive the one-loop axion couplings to charged leptons under two assumptions.
Finally, Sec.~\ref{sec:conc} summarizes our main results.
\ref{ap:proof} provides the proof of the statements on the PQ charge assignment presented in Sec.~\ref{subsec:model}.
\ref{ap:num} presents the numerical analysis of the neutrino mass texture, while \ref{ap:clock} discusses the clockwork enhancement which may be used to enhance the axion--neutrino coupling.
\ref{ap:calc} lists the loop integrals used in the calculation of the axion--charged lepton couplings and \ref{ap:id} presents the proof of the matrix equalities used in the cancellation of the ultraviolet (UV) divergences.

\section{Model building and the texture of $M_\nu$}
\label{sec:model}

\subsection{Model building}\label{subsec:model}

In this section, we combine the KSVZ model with the Type-I seesaw mechanism \cite{Frampton2002,Ibarra2004,Harigaya2012} to construct a minimal flavored $\nu$KSVZ model. 

Following the KSVZ framework, we introduce a electroweak singlet heavy quark field $Q$, which couples chirally to scalar fields $\phi_i$ in order to solve the strong CP problem. At the same time, these scalar fields provide Majorana masses for the three generations of right-handed neutrinos $N_{1R}, N_{2R}, N_{3R}$ through Yukawa interactions 
\begin{equation}
	x_{ijk}\overline{N^c_{iR}} N_{jR}\phi_k^{(*)},
\end{equation}
where $\phi^{(*)}_k$ stands for either $\phi_k$ or its complex conjugate $\phi_k^*$.

We assign PQ charges $\chi_1$, $\chi_2$, and $\chi_3$ to the right-handed neutrinos $N_{1R}$, $N_{2R}$, and $N_{3R}$, respectively. This assignment effectively induces PQ charges for the Yukawa coefficients $x_{ijk}$. The PQ charge of each coefficient, corresponding to the bilinear operator $\overline{N^c_{iR}} N_{jR}$, is given by
\begin{equation}
	\chi(\overline{N^c_{iR}} N_{jR})=\begin{pmatrix}
		2\chi_1&\chi_1+\chi_2&\chi_1+\chi_3\\
		\chi_1+\chi_2&2\chi_2&\chi_2+\chi_3\\
		\chi_1+\chi_3&\chi_2+\chi_3&2\chi_3
	\end{pmatrix}.
\end{equation}
PQ symmetry invariance requires that each coefficient $x_{ijk}$ must either vanish or be compensated by a scalar field $\phi_k$ carrying the opposite PQ charge.

We define $N_{PQ}$ as the number of distinct PQ charges present in the Yukawa coefficients $x_{ijk}$. This allows us to systematically classify the PQ charge assignments as follows:
\begin{itemize}
	\item[1.] $N_{PQ}=1$:\\
	This corresponds to the flavor-universal scenario, which requires only one single scalar field.
	\item[2.] $N_{PQ}=2$: \\
	It is straightforward to verify that this condition is algebraically impossible to realize.
	\item[3.] $N_{PQ}=3$: \\
	This is the minimal scenario that allows for a nontrivial flavor structure.
	Up to the permutation of indices $(1,2,3)$ for the right-handed neutrinos, there is a unique assignment:
	\begin{equation}
			\chi(\overline{N^c_{iR}} N_{jR})=\begin{pmatrix}
			2\chi_1&\chi_1+\chi_2&\chi_1+\chi_2\\
			\chi_1+\chi_2&2\chi_2&2\chi_2\\
			\chi_1+\chi_2&2\chi_2&2\chi_2
		\end{pmatrix},
	\end{equation}
	where $\chi_1\neq \chi_2$.
\end{itemize}
The detailed proof that the $N_{PQ}=2$ scenario is impossible, together with the uniqueness (up to permutations) of the $N_{PQ}=3$ assignment, is presented in \ref{ap:proof}.

While $N_{PQ} > 3$ is possible, we focus on the $N_{PQ}=3$ scenario as the minimal flavored construction. To further refine the model, we impose a $\mu-\tau$ symmetry on the charge assignments, which is motivated by neutrino phenomenology and effectively reduces the remaining permutation redundancies.
In this minimal scenario, ensuring matrix invertibility and non-vanishing mixing angles requires at least two scalars, $\phi_1$ and $\phi_2$, with distinct PQ charges. When only two scalars are introduced, the structure of the Yukawa matrix is uniquely fixed as:
\begin{equation}
	 x_{ijk}\phi_k^{(*)}=\begin{pmatrix}
		0&x_{121}\phi_1&x_{131}\phi_1\\
		x_{121}\phi_1&x_{222}\phi_2&x_{232}\phi_2\\
		x_{131}\phi_1&x_{232}\phi_2&x_{332}\phi_2
	\end{pmatrix}.
\end{equation}
Note that a special case exists where $\phi_2 = \phi_1^*$, effectively reducing the sector to a single scalar field. 
However, this configuration suffers from an unsatisfactory drawback: to account for the observed neutrino masses and mixing angles, one must assume a specific hierarchy among the Yukawa couplings $x_{ijk}$, which leads to a ``fine-tuning'' problem (see the discussion in Sec.~\ref{sec:VEV}). Moreover, this configuration fixes the PQ charge assignments to $(\chi_1,\chi_2,\chi_3)=(1,-1/3,-1/3)$ up to an overall normalization, thereby making it impossible to enhance the axion--neutrino couplings by enlarging the differences among the $\chi_i$'s (see \ref{ap:clock} for a discussion).

By requiring minimality in both $N_{PQ}$ and the number of additional scalars, the PQ charge assignment is uniquely determined. We summarize these assignments in Table~\ref{tab:1}, normalizing to $\chi_1=1$. The only free parameter is the charge difference between the first and the latter two generations (i.e., $\chi - 1$), which must be non-zero.
\begin{table}[h]
	\centering
	\begin{tabular}{c|c|c|c|c|c|c}
		& $Q$ & $f_1$ & $f_2$ & $f_3$ & $\phi_1$ & $\phi_2$\\
		\hline
		PQ charge & $\gamma_5 q$ & 1 & $\chi$ & $\chi$ & $-1-\chi$ & $-2\chi$
	\end{tabular}
	\caption{The PQ charge assignments in this model. Here $\chi\neq 1$, and $f_i$ denotes $N_{iR}$, $l_{iR}$, and $L_i$. If $Q$ couples to $\phi_1$ in the Lagrangian, then $q=(1+\chi)/2$, while if $Q$ couples to $\phi_2$, then $q=\chi$.}
	\label{tab:1}
\end{table}

The most general scalar potential \cite{DILUZIO20201} takes the form
\begin{align}
	V(H,\phi_1,\phi_2)=&\,\tilde{V}_{\rm moduli}(|H|,|\phi_1|,|\phi_2|)
	+\tilde{V}_{\rm break}(H,\phi_1,\phi_2)+{\rm h.c.}.\label{eq:potential}
\end{align}
The first term contains all moduli terms allowed by gauge invariance. The second term explicitly breaks the two independent rephasing symmetries of the SM-singlet scalars down to a single $U(1)_{PQ}$,
\begin{align}
	U(1)_{\phi_1}\times U(1)_{\phi_2}\rightarrow U(1)_{PQ},
\end{align}
thereby ensuring that the low-energy spectrum contains only a single light axion.
For example, such a term can be chosen as
\begin{equation}
	\tilde{V}_{\rm break}(H,\phi_1,\phi_2)=\alpha\phi_1^n\phi_2^m+h.c.,\quad n,m\in\mathbb{Z}^+,\label{eq:Vb}
\end{equation}
and the PQ symmetry conservation requires that
\begin{equation}
	n\chi_{\phi_1}+m\chi_{\phi_2}=0,\label{eq:nm}
\end{equation}
where $\chi_{\phi_1}=-1-\chi$ and $\chi_{\phi_2}=-2\chi$ are the PQ charges of $\phi_1$ and $\phi_2$. For other choices of the explicit breaking term, the constraint in Eq.~\eqref{eq:nm} should be modified accordingly. For example, if $\phi_1$ in Eq.~\eqref{eq:Vb} is replaced by $\phi_1^*$, the integer $n$ in Eq.~\eqref{eq:nm} is simply replaced by $-n$.
If $n+m\le4$, the explicit breaking term is renormalizable; otherwise, it is non-renormalizable. Throughout this work, we assume that the explicit breaking is sufficiently strong that the phase field orthogonal to the axion is much heavier than the axion and can therefore be integrated out.
	
By an appropriate choice of the scalar potential in Eq.~\eqref{eq:potential}, the three scalar fields can always acquire the following VEVs:
\begin{align}
	\langle H\rangle=\frac{\lambda_H}{\sqrt{2}} 
	\begin{pmatrix}
		0\\1
	\end{pmatrix},\quad  
	\langle\phi_{1,2}\rangle=\frac{\lambda_{1,2}}{\sqrt{2}}.\label{eq:VEV}
\end{align}
With these VEVs, the scalar fields can be parametrized as 
\begin{equation}
	H=\begin{pmatrix}
		G^+\\
		\frac{1}{\sqrt{2}}(\lambda_H+h+\mathrm{i}G^0)
	\end{pmatrix},\quad 
	\phi_{1,2}=\frac{1}{\sqrt{2}}(\lambda_{1,2}+\rho_{1,2})\mathrm{e}^{\mathrm{i}a_{1,2}/\lambda_{1,2}} .
\end{equation}
The relevant Yukawa interactions are
\begin{align}
	\mathcal{L}_Y\supset&\, x_{ijk}\overline{N^c_{iR}} N_{jR}\phi_k 
	+y^{(D)}_{ij}\bar{L}_i N_{jR}\tilde{H}
	+y^{(l)}_{ij}\bar{L}_i l_{jR}H+{\rm h.c.},\label{eq:Yukawa}
\end{align}
and, after inserting the VEVs in Eq.~\eqref{eq:VEV}, the resulting mass matrices take the form
\begin{align}
	M^{(R)}&=\frac{1}{\sqrt{2}}
	\begin{pmatrix}
		0&y^{(R)}_{12}\lambda_1&y^{(R)}_{13}\lambda_1\\
		y^{(R)}_{12}\lambda_1&y^{(R)}_{22}\lambda_2&y^{(R)}_{23}\lambda_2\\
		y^{(R)}_{13}\lambda_1&y^{(R)}_{23}\lambda_2&y^{(R)}_{33}\lambda_2
	\end{pmatrix},\label{eq:MR}\\
	M^{(D)}&=\frac{\lambda_H}{\sqrt{2}}
	\begin{pmatrix}
		y^{(D)}_{11}&0&0\\
		0&y^{(D)}_{22}&y^{(D)}_{23}\\
		0&y^{(D)}_{32}&y^{(D)}_{33}
	\end{pmatrix},\label{eq:MD}\\
	M^{(l)}&=\frac{\lambda_H}{\sqrt{2}}
	\begin{pmatrix}
		y^{(l)}_{11}&0&0\\
		0&y^{(l)}_{22}&y^{(l)}_{23}\\
		0&y^{(l)}_{32}&y^{(l)}_{33}
	\end{pmatrix}.\label{eq:Ml}
\end{align}
Here $y_{12}^{(R)}=x_{121}$, $y_{13}^{(R)}=x_{131}$, and similarly for the other entries.

\subsection{The texture of $ M_\nu $}
\label{sec:texture1}
	
As we will see, this model predicts a specific texture for the neutrino mass matrix. In the canonical Type-I seesaw framework, we assume $M^{(D)}\ll M^{(R)}$, so that \cite{cai2018lepton}
\begin{align}
	\tilde{M}_\nu=-M^{(D)}{M^{(R)}}^{-1}{M^{(D)}}^T,\label{eq:seesaw1}
\end{align}
where $\tilde{M}_\nu$ denotes the mass matrix of the light neutrinos in the original basis.

Assuming that there is no massless neutrino, we can take the inverse of both sides of Eq.~\eqref{eq:seesaw1}, which yields
\begin{align}
	M^{(R)}=-{M^{(D)}}^T\tilde{M}_\nu^{-1}M^{(D)}.\label{eq:seesaw2}
\end{align}
Notice that the zeros in the first row and column of $M^{(D)}$ isolate the $(1,1)$ element of $\tilde{M}_\nu^{-1}$. So the zero $(1,1)$ element of $M^{(R)}$ implies 
\begin{align}
	(\tilde{M}_\nu^{-1})_{11}=0.\label{eq:texture0}
\end{align}
This is a zero texture of $\tilde{M}_\nu^{-1}$. It can equivalently be written as
\begin{equation}
	(\tilde{M}_\nu)_{22}(\tilde{M}_\nu)_{33}=(\tilde{M}_\nu)_{23}^2.\label{eq:texture00}
\end{equation}
This relation can also be derived directly from Eq.~\eqref{eq:seesaw1} using the adjugate-matrix technique, even when $\tilde{M}_\nu$ is not invertible.
	
Since the charged-lepton mass matrix in Eq.~\eqref{eq:Ml} is block diagonal, it can be diagonalized as
\begin{align}
	\hat{M}^{(l)}=U_{lL}^\dagger M^{(l)}U_{lR},
\end{align}
with block-diagonal unitary matrices
\begin{align}
	U_{lL}=\begin{pmatrix}
		u_L&0\\
		0&\tilde{U}_{lL}
	\end{pmatrix},\quad 
	U_{lR}=\begin{pmatrix}
		u_R&0\\
		0&\tilde{U}_{lR}
	\end{pmatrix}.\label{eq:UlL}
\end{align}
Here $u_L$ and $u_R$ are complex numbers with unit modulus, and $\tilde{U}_{lL}$ and $\tilde{U}_{lR}$ are $2\times2$ unitary matrices.

Using the left-handed charged-lepton rotation matrix $U_{lL}$, the light-neutrino mass matrix in the neutrino flavor basis is given by \cite{Bj_rkeroth_2020}
\begin{align}
	M_\nu=U_{lL}^\dagger \tilde{M}_\nu U_{lL}^*.\label{eq:MnuMtilde}
\end{align}
From Eqs.~\eqref{eq:texture0} and \eqref{eq:UlL}, we see that the same texture also holds for $M_\nu$, since $U_{lL}$ is unitary within each block. In particular,
\begin{align}
	(M_\nu^{-1})_{11}=0 
	\quad\text{and}\quad 
	(M_\nu)_{22}(M_\nu)_{33}=(M_\nu)_{23}^2.\label{eq:texture1}
\end{align}
This texture is a direct consequence of this minimal model.

While Ref. \cite{Lashin2009} provides a comprehensive numerical survey of this texture, an approximate analytical treatment is essential to elucidate the physical mechanism driving the IH prediction. In the following section, we derive the explicit relations between the Majorana phases, $m_3$, and $\langle m_{\beta\beta}\rangle$. This approach ensures the inclusion of some configurations--such as the $\eta_1 = \eta_2$ case--which may be undersampled or overlooked in purely numerical scans due to finite resolution.

\section{Inverted mass ordering and small $\theta_{13}$}
\label{sec:hierarchy}

\subsection{Inverted neutrino mass hierarchy}
\label{sec:IH}

In this section, we show that the texture $(M_\nu^{-1})_{11}=0$ is compatible with current experimental data and leads to an inverted mass ordering, while establishing definite relations among the PMNS parameters and neutrino masses.

The neutrino mass matrix $M_\nu$ can be diagonalized by the PMNS matrix \cite{nomura2023texture} as
\begin{align}
	M_\nu=U \hat{M}_\nu U^T,\label{eq:UMU}
\end{align}
where
\begin{align}
	\hat{M}_\nu=\mathrm{diag}(m_1,m_2,m_3),
\end{align}
and $U$ denotes the PMNS matrix. We adopt the following parametrization \cite{PDG2024}:
\begin{align}
	U=VP,\label{eq:UVP}
\end{align}
with
\begin{align}
	&V=\begin{pmatrix}
		c_{12} c_{13} & s_{12} c_{13} & s_{13} \mathrm{e}^{-\mathrm{i} \delta_{13}} \\
		-s_{12} c_{23}-c_{12} s_{23} s_{13} \mathrm{e}^{\mathrm{i} \delta_{13}} & c_{12} c_{23}-s_{12} s_{23} s_{13} \mathrm{e}^{\mathrm{i} \delta_{13}} & s_{23} c_{13} \\
		s_{12} s_{23}-c_{12} c_{23} s_{13} \mathrm{e}^{\mathrm{i} \delta_{13}} & -c_{12} s_{23}-s_{12} c_{23} s_{13} \mathrm{e}^{\mathrm{i} \delta_{13}} & c_{23} c_{13}
	\end{pmatrix},\quad
	P=\begin{pmatrix}
		\mathrm{e}^{\mathrm{i}\eta_1}&0&0\\
		0&\mathrm{e}^{\mathrm{i}\eta_2}&0\\
		0&0&1
	\end{pmatrix},
\end{align}
where $s_{ij}\equiv \sin\theta_{ij}$ and $c_{ij}\equiv\cos\theta_{ij}$.

Following a procedure similar to that of Refs.~\cite{fritzsch2011two,PhysRevD.107.035017}, we insert Eq.~\eqref{eq:UMU} into the texture condition~\eqref{eq:texture1} and obtain
	\begin{align}
		m_1 m_2 s_{13}^2 
		+ m_1 m_3 s_{12}^2 c_{13}^2 \mathrm{e}^{-2\mathrm{i}(\eta_2+\delta_{13})}
		+ m_2 m_3 c_{12}^2 c_{13}^2 \mathrm{e}^{-2\mathrm{i}(\eta_1+\delta_{13})}
		=0.\label{eq:texture3}
	\end{align}
This is a complex equation, corresponding to two real constraints. Since $\Delta m^2_{21}$, $\Delta m^2_{32}$ and the mixing angles $\theta_{12}$, $\theta_{23}$, and $\theta_{13}$ are already known from experiments, these two equations allow one, in principle, to solve for the neutrino masses $m_i$ and the phase combination $\eta_1+\delta_{13}$ once $\eta_2+\delta_{13}$ is specified. Also, this equation implies that there is no massless neutrino.

From a geometrical point of view, the condition for Eq.~\eqref{eq:texture3} to vanish is equivalent to the requirement that the three terms on the left-hand side form a closed triangle in the complex plane. The existence of such a triangle is governed by the triangle inequality, which yields the following necessary condition:
\begin{equation}
	m_1 m_2 s_{13}^2 \geq m_2 m_3 c_{12}^2 c_{13}^2 - m_1 m_3 s_{12}^2 c_{13}^2.
\end{equation}
Given the experimental ordering $m_1 < m_2$, this inequality leads to a lower bound on the mass ratio:
\begin{equation}
	\frac{m_1}{m_3} > \frac{c_{12}^2 - s_{12}^2}{t_{13}^2},
\end{equation}
where $t_{13} \equiv \tan\theta_{13}$. Adopting the central values from the current global fit data \cite{PDG2024}, one finds $(c_{12}^2 - s_{12}^2)/t_{13}^2 \approx 17$. This immediately implies that the texture condition in Eq.~\eqref{eq:texture3} can only be satisfied if the neutrinos follow the IH \cite{neutrino_physics}, characterized by:
\begin{equation}
	m_3 \ll m_1 \simeq m_2.\label{eq:IH1}
\end{equation}
By rescaling Eq.~\eqref{eq:texture3} by $m_3$ and invoking the quasi-degenerate limit $m_1 \simeq m_2$, the texture condition reduces to:
\begin{align}
	\frac{m_1}{m_3}
	=-\frac{1}{t_{13}^2}
	\Big[s_{12}^2\,\mathrm{e}^{-2\mathrm{i}(\eta_2+\delta_{13})}
	+c_{12}^2\,\mathrm{e}^{-2\mathrm{i}(\eta_1+\delta_{13})}\Big].
	\label{eq:IH2}
\end{align}
We first examine the solvability of this relation. Given that the Majorana phases $\eta_{1,2}$ in Eq.~\eqref{eq:IH2} possess a period of $\pi$, we can, without loss of generality, restrict their ranges to the interval $[0, \pi)$.

To ensure a physical solution for Eq.~\eqref{eq:IH2}, the imaginary part of its right-hand side must vanish. This consistency condition yields a direct relation between the two Majorana phases $\eta_1$ and $\eta_2$:
	\begin{equation}
		\sin 2(\eta_1+\delta_{13})
		=-t_{12}^2\,
		\sin 2(\eta_2+\delta_{13}).
		\label{eq:IH3}
	\end{equation}
where $t_{12} \equiv \tan\theta_{12}$. Experimental data indicates $t^2_{12} \approx 0.44$, ensuring that the magnitude of the right-hand side is always bounded by unity. Consequently, for any given $\eta_2$, Eq.~\eqref{eq:IH3} consistently admits two distinct solutions for $\eta_1$ within the $[0, \pi)$ interval, denoted as $\eta_1$ and $\eta_1'$, which are related by:
	\begin{align}
		(\eta_1+\delta_{13})+(\eta_1'+\delta_{13})
		=\frac{\pi}{2}+n\pi,\qquad n\in\mathbb{Z}.
		\label{eq:eta}
	\end{align}
The remaining requirement is that the real part of the right-hand side of Eq~\eqref{eq:IH2} must be positive to ensure $m_1/m_3 > 0$, which leads to the condition:
\begin{equation}
	s_{12}^2\cos 2(\eta_2+\delta_{13})
	+ c_{12}^2\cos 2(\eta_1+\delta_{13}) < 0 .
	\label{eq:IH4}
\end{equation}
To demonstrate that one of the two solutions $\{\eta_1, \eta_1'\}$ always satisfies this requirement, we note that Eq.~\eqref{eq:IH3} further implies the auxiliary inequality:
\begin{align}
	c_{12}^2\big|\cos 2(\eta_1+\delta_{13})\big|
	> s_{12}^2\big|\cos 2(\eta_2+\delta_{13})\big| .
\end{align}
By virtue of this relation, if a given $\eta_1$ fails to satisfy Eq.~\eqref{eq:IH4}, its complementary solution $\eta_1'$ (defined in Eq.~\eqref{eq:eta}) will necessarily fulfill it.

Consequently, for any given $\eta_2$, Eq.~\eqref{eq:IH2} always admits a unique physical solution for the pair $(m_3, \eta_1)$ that preserves the positivity of the mass eigenvalues.

It is noteworthy that Eq.~\eqref{eq:IH3} admits two special configurations for the phases:
\begin{align}
	&1) \eta_1+\delta_{13}=\eta_2+\delta_{13}=\frac{\pi}{2},\label{eq:eta1=eta2}\\
	&2)\eta_1+\delta_{13}=\frac{\pi}{2},\; \eta_2+\delta_{13}=0.\label{eq:eta1=eta2+pi/2}
\end{align}
These two cases are independent of the specific values of the mixing parameters. Furthermore, by substituting these points back into the exact texture condition in Eq.~\eqref{eq:texture3}, one can verify that they represent exact solutions corresponding to the minimum and maximum values of $m_3$.

Next, we derive the neutrino mass ratio from Eq. \eqref{eq:IH2}. It is evident that the order of magnitude for this ratio is governed by:
\begin{equation}
	\frac{m_3}{m_1} \sim s_{13}^2 ,
	\label{eq:GST}
\end{equation}
which can be interpreted as an approximate (inverted) GST-like relation \cite{Gatto1968,Fritzsch1977,Roy2020}.
However, the precise magnitude of the ratio is sensitive to the specific values of the Majorana phases. Specifically, the following bound holds:
\begin{align}
	\big|c_{12}^2 - s_{12}^2\big|
	\leq \big| s_{12}^2 \mathrm{e}^{-2\mathrm{i}(\eta_2+\delta_{13})}
	+ c_{12}^2 \mathrm{e}^{-2\mathrm{i}(\eta_1+\delta_{13})} \big|
	\leq 1 ,\label{eq:c12-s12}
\end{align}
where the lower and upper bounds are saturated by the phase configurations identified in Eq.~\eqref{eq:eta1=eta2+pi/2}  and Eq.~\eqref{eq:eta1=eta2}, respectively.

By adopting the central values of the mixing parameters from Ref. \cite{PDG2024}, we obtain the following permissible range for the mass ratio:
\begin{align}
	17 \lesssim \frac{m_1}{m_3} \lesssim 45 ,
	\label{eq:massratio}
\end{align}
which constrains the lightest neutrino mass to the interval:
\begin{align}
		1.1~\mathrm{meV} \lesssim m_3 \lesssim 3.0~\mathrm{meV},
\end{align} 
and the total neutrino mass $\Sigma=m_1+m_2+m_3$ falls within
\begin{equation}
	0.101~\mathrm{eV}\lesssim\Sigma\lesssim0.103~\mathrm{eV}.
\end{equation}

Finally, we evaluate the effective Majorana mass, which is defined generally as:
\begin{equation}
		\langle m_{\beta\beta}\rangle=|\sum_{i}(U_{1i})^2m_i|.
\end{equation}
In the IH regime, by invoking the quasi-degeneracy $m_1 \simeq m_2$, this general expression can be simplified to:
	\begin{equation}
		\langle m_{\beta\beta}\rangle=m_1 c_{13}^2|s^2_{12}\mathrm{e}^{-2\mathrm{i}(\eta_2+\delta_{13})}+c^2_{12}\mathrm{e}^{-2\mathrm{i}(\eta_1+\delta_{13})}|.\label{eq:mbeta}
\end{equation}
Notably, the structural form of Eq.~\eqref{eq:mbeta} is highly analogous to that of the mass ratio in Eq.~\eqref{eq:IH2}, implying that the magnitude of $\langle m_{\beta\beta}\rangle$ is similarly dictated by the interference of the Majorana phases. By combining Eq.~\eqref{eq:IH2} with Eq.~\eqref{eq:mbeta}, we obtain a remarkably simple approximate analytical relation:
\begin{equation}
	\langle m_{\beta\beta}\rangle=\frac{m_1^2}{m_3}s_{13}^2.
\end{equation}
Adopting the central values of the parameters from Ref.~\cite{PDG2024}, this yields the following predicted range for the effective Majorana mass:
\begin{equation}
	19~\mathrm{meV}\lesssim\langle m_{\beta\beta}\rangle\lesssim 50~\mathrm{meV}. \label{eq:mbetabeta}
\end{equation}

Summary. In short, within the IH scenario Eq.~\eqref{eq:texture3} consistently admits valid solutions for the neutrino masses $m_i$ and the Majorana phase $\eta_1$ for any prescribed value of $\eta_2$. Both $m_3$ and $\langle m_{\beta\beta}\rangle$ are sensitive to the relative Majorana phases; specifically, the configurations in Eq.~\eqref{eq:eta1=eta2} (Eq.~\eqref{eq:eta1=eta2+pi/2}) correspond to the minimum (maximum) value of $m_3$ and the maximum (minimum) value of $\langle m_{\beta\beta}\rangle$.
We numerically verify these analytical results in \ref{ap:num} by directly solving Eq.~\eqref{eq:texture3}, taking into account both the central values and the associated uncertainties of the experimental data. 

\subsection{Small $s_{13}$ and the hierarchy of the two scalar VEVs}
\label{sec:VEV}

From Eq.~\eqref{eq:GST}, we see that within the texture of this model, a small $s_{13}$ is correlated with the neutrino mass hierarchy. In fact, we will show that the smallness of $s_{13}$ mainly originates from the hierarchy between the VEVs of the two scalar fields in this model.

The relation between the light neutrino mass matrix in the original basis and that in the mass basis is
\begin{align}
	\hat{M}_\nu = U_\nu^\dagger \tilde{M}_\nu U_\nu^*, \label{eq:hierM0}
\end{align}
where $U_\nu = U_{lL} U$. Hence, combining Eq.~\eqref{eq:seesaw2} with Eq.~\eqref{eq:hierM0}, we obtain the relation between $M^{(R)}$ and $\hat{M}_\nu$,
\begin{align}
	M^{(R)} = - {M^{(D)}}^T U_\nu^* \hat{M}_\nu^{-1} U_\nu^\dagger M^{(D)}. \label{eq:hierM}
\end{align}

If we keep only the dominant terms of $\mathcal{O}(1/m_3)$, since $m_3 \ll m_1, m_2$, the entries of $M^{(R)}$ can be approximated as\footnote{
	For Eq.~\eqref{eq:MR13}, retaining only the leading $\mathcal{O}(1/m_3)$ contribution results in an approximation error of about $3\%$. For Eq.~\eqref{eq:MR12}, the leading term is proportional to the small matrix element $(U_\nu^\dagger)_{31}$, so the relative error may be moderately enhanced, but typically remains at the several-percent level.
}
\begin{align}
	M^{(R)}_{12} &\simeq -\frac{\alpha}{m_3} M^{(D)}_{11} (U^\dagger_\nu)_{31}, 
	\qquad 
	M^{(R)}_{13} \simeq -\frac{\beta}{m_3} M^{(D)}_{11} (U^\dagger_\nu)_{31},\label{eq:MR12} \\
	M^{(R)}_{22} &\simeq -\frac{\alpha^2}{m_3}, 
	\qquad 
	M^{(R)}_{23} \simeq -\frac{\alpha \beta}{m_3}, 
	\qquad 
	M^{(R)}_{33} \simeq -\frac{\beta^2}{m_3},\label{eq:MR13}
\end{align}
where
\begin{align}
	\alpha &= (U_\nu^\dagger)_{32} M^{(D)}_{22} + (U_\nu^\dagger)_{33} M^{(D)}_{32}, \\
	\beta  &= (U_\nu^\dagger)_{32} M^{(D)}_{23} + (U_\nu^\dagger)_{33} M^{(D)}_{33}.
\end{align}
Since $|(U^\dagger_\nu)_{31}|$ is small \cite{PDG2024},
\begin{equation}
	|(U^\dagger_\nu)_{31}| = |u_L s_{13} \mathrm{e}^{-\mathrm{i}\delta_{13}}| = s_{13} \approx 0.148,
\end{equation}
we find that the matrix elements of $M^{(R)}$ exhibit the structure
\begin{align}
	\frac{|M^{(R)}_{12,13}|}{|M^{(R)}_{22,23,33}|} \sim |(U^\dagger_\nu)_{31}| = s_{13}.
\end{align}

On the other hand, from the structure of the $M^{(R)}$ matrix in Eq.~\eqref{eq:MR}, and assuming that the relevant Majorana Yukawa couplings do not exhibit a strong hierarchy, we naturally obtain
\begin{align}
	\frac{|M^{(R)}_{12,13}|}{|M^{(R)}_{22,23,33}|}
	\sim
	\frac{\lambda_1}{\lambda_2}.
\end{align}
Under this assumption, the model naturally leads to the relation
\begin{equation}
	s_{13} \sim \frac{\lambda_1}{\lambda_2}.
\end{equation}
In this model, the presence of two scalar fields does not require their VEVs to lie at the same energy scale. When a hierarchy develops between these VEVs, it naturally explains the smallness of $s_{13}$.
The numerical examination of the hierarchy in $M^{(R)}$ is presented in \ref{ap:num}.

\subsection{Phenomenological implications in the neutrino sector}
Before proceeding to the discussion of the axion sector, we summarize the phenomenological implications of the neutrino sector and discuss the corresponding experimental constraints.

The neutrino mass ratio in Eq.~\eqref{eq:massratio} leads to the following phenomenological predictions:
\begin{itemize}
	\item[(i)] The neutrino mass ordering is inverted ordering (IO).
	\item[(ii)] The total neutrino mass is approximately $0.101~\text{eV}\text{--}0.103~\text{eV}$.
	\item[(iii)] The effective Majorana mass lies in the range $19~\mathrm{meV}\text{--} 50~\mathrm{meV}$.
\end{itemize}
Among these predictions, the most distinctive one is the inverted neutrino mass ordering. Current global oscillation analyses show a mild preference for normal ordering (NO) at the level of approximately $2.2$--$2.5\sigma$ \cite{Esteban2024,Capozzi2025,JUNO2025,Esteban2026}. Accordingly, the model is currently disfavored, but not excluded, by oscillation data. It would be falsified if future oscillation experiments establish the NO with high significance (e.g. at the $5\sigma$ level). Furthermore, establishing the NO would not only falsify the present model, but also exclude a broader class of models that predict the same $(M_\nu^{-1})_{11}=0$ texture or are based on the same PQ charge assignment. Conversely, although currently less favored by the data, it remains possible that future oscillation experiments will establish the IO with high significance. In that case, the present model would remain viable, and its predictions in the axion sector would provide complementary experimental tests.

The next important constraint comes from cosmology. Recent analyses combining DESI BAO measurements with CMB data from the Planck satellite and the Atacama Cosmology Telescope (ACT) set a particularly stringent limit, $\Sigma<0.072~\text{eV}$ at $95\%$ CL \cite{Adame2025}.
This bound lies below the minimum neutrino mass sum allowed by oscillation data for the IO and is close to the minimum value for the NO. Nevertheless, we note that the robustness of this limit has been questioned in the recent literature~\cite{Jiang2025,Capozzi2025}. In particular, within the standard $\Lambda$CDM framework, there exists a tension between the DESI BAO data and terrestrial neutrino measurements. First, the tightest analyses within the standard $\Lambda$CDM framework further strengthen the upper limit to $\Sigma<0.05~\text{eV}$ \cite{Jiang2025}, which is already in tension with the oscillation constraint $\Sigma\gtrsim0.06~\text{eV}$. Second, if the physical prior $\Sigma>0$ is relaxed and the likelihood is allowed to extend into the unphysical region $\Sigma<0$, the best-fit value is shifted to negative neutrino masses. These observations suggest that the most stringent cosmological limits should be interpreted with some caution.
Following the discussion in Ref.~\cite{Capozzi2025}, we adopt the conservative summary of the present cosmological constraints,
\begin{equation}
	\Sigma<0.2~\text{eV}\;\text{at}\;2\sigma\;\text{(within a factor of 3)},
\end{equation}
which reflects the spread of current upper limits arising from different data combinations and from extensions of the standard $\Lambda$CDM framework. Under this more conservative interpretation, the prediction of the present model, $\Sigma\simeq0.10~\text{eV}$
remains viable under current cosmological observations, although it is in significant tension with the most stringent limit obtained within the standard $\Lambda$CDM framework.

Another important constraint comes from neutrinoless double beta decay ($0\nu\beta\beta$) experiments. At present, the KamLAND-Zen experiment \cite{Abe2025} gives the most stringent constraint on $\langle m_{\beta\beta}\rangle$:
\begin{equation}
	\langle m_{\beta\beta}\rangle<(28-122)~\mathrm{meV}\quad (\text{$90\%$ CL}),
\end{equation}
where the range of the upper bound reflects the uncertainties in various nuclear matrix element (NME) calculations. This interval partially overlaps with the predicted range of $\langle m_{\beta\beta}\rangle$. Nevertheless, even if the most stringent NME is adopted, i.e. $\langle m_{\beta\beta}\rangle<28~\mathrm{meV}$, a substantial fraction of the predicted parameter space (see Fig.~\ref{fig:mee}) remains allowed.
However, the projected sensitivities of several next-generation $0\nu\beta\beta$ experiments, including LEGEND-1000, CDEX-1T, and CUPID, indicate that the entire predicted range of $\langle m_{\beta\beta}\rangle$ in this model could be tested, although the exact sensitivities depend on the adopted experimental assumptions and NME calculations \cite{Yu2025,LEGEND2021,CUPID2022}.

Taken together, these considerations show that the present model can be decisively tested in the near future. It would be falsified if future oscillation experiments establish the normal ordering with high significance, if cosmological observations robustly exclude $\Sigma\simeq0.10~\mathrm{eV}$, or if next-generation $0\nu\beta\beta$ experiments exclude the predicted interval of $\langle m_{\beta\beta}\rangle$. Such a result would also have implications beyond the present model, excluding a broader class of models that predict the same $(M_\nu^{-1})_{11}=0$ texture or are constructed from the same PQ charge assignment.

\section{Phenomenology of axion--lepton couplings}
\label{sec:couplings}
In this section, we investigate the axion--lepton interactions in the flavored $\nu$KSVZ model and highlight the features that distinguish them from those in flavor-universal models.
We first identify the axion field in this framework and then derive the leading-order effective interactions between the axion and leptons. Beyond the standard KSVZ framework, our model predicts additional off-diagonal couplings. The corresponding Lagrangian and its phenomenological implications are discussed below.

\subsection{Identification of the axion}	
From the PQ charges of the scalar fields, the axion field, corresponding to the Goldstone mode associated with the spontaneous breaking of the PQ symmetry, is given by
\begin{equation}
	a=\frac{1}{\lambda}(\lambda_1\chi_{\phi_1}a_1+\lambda_2\chi_{\phi_2} a_2),\quad 
	\lambda=\sqrt{\lambda_1^2\chi_{\phi_1}^2+\lambda_2^2\chi_{\phi_2}^2},
\end{equation}
together with the orthogonal combination
\begin{equation}
	b=\frac{1}{\lambda}(\lambda_2\chi_{\phi_2}a_1-\lambda_1\chi_{\phi_1} a_2).  
\end{equation}
Since the contribution to the mass from the explicit breaking term is assumed to be much larger than that induced by QCD instantons, both $a$ and $b$ are approximate mass eigenstates~\cite{Kim2005,Choi2014}. The corresponding approximation is accurate up to corrections of order $m_a^2/m_b^2$.

For the electroweak singlet heavy quark $Q$, the Yukawa interaction can involve either $\phi_1^{(*)}$ or $\phi_2^{(*)}$. For illustration, we consider the case
\begin{equation}
	Y^{(Q)}\bar{Q}_LQ_R\phi_1+h.c.,
\end{equation}
and the generalization to the alternative choice is straightforward.

After the field-dependent transformation
\begin{equation}
	Q\rightarrow\exp\Big(-\mathrm{i}\frac{\chi_{\phi_1}a}{2\lambda}\gamma_5\Big)Q,
\end{equation}
the corresponding color anomaly term is
\begin{equation}
	\frac{\chi_{\phi_1}a}{\lambda}\frac{g_s^2}{32\pi^2}G\tilde{G},\label{eq:anomaly}
\end{equation}
where $g_s$ is the gauge coupling of $SU(3)_C$. Then the decay constant of the axion $f_a$, which is defined by normalizing the coefficient of the QCD anomaly \cite{DILUZIO20201,Cox2024}, is 
\begin{equation}
	f_a=\lambda/\chi_{\phi_1}.
\end{equation}
The condition \eqref{eq:nm} implies that the ratio $\chi_{\phi_2}/\chi_{\phi_1}$ is rational. Therefore, both $\chi_{\phi_1}$ and $\chi_{\phi_2}$ can always be rescaled to integers simultaneously. Denote the PQ charges of scalars after scaling as
\begin{equation}
	\chi_{\phi_1}'=s\chi_{\phi_1},\quad \chi_{\phi_2}'=s\chi_{\phi_2}.
\end{equation}
This rescaling can be made unique (up to an overall sign) by requiring 
\begin{equation}
	\operatorname{gcd}(\chi_{\phi_1}',\chi_{\phi_2}')=1,
\end{equation}
where $\operatorname{gcd}(\chi_{\phi_1}',\chi_{\phi_2}')$ is the greatest common divisor of $\chi_{\phi_1}'$ and $\chi_{\phi_2}'$.

With this convention, the domain wall number of the axion is \cite{Ernst2018,Diehl2023}
\begin{equation}
	N_{DW}=|\chi_{\phi_1}'|.
\end{equation}
If $N_{DW}>1$ and the spontaneous PQ symmetry breaking occurs after inflation, the resulting stable domain walls lead to the well-known domain wall problem. In this case, additional mechanisms, such as introducing a bias term to lift the vacuum degeneracy, are required to eliminate the domain walls~\cite{Sikivie1982}.

\subsection{The axion--neutrino couplings}\label{sec:AxionNu}
In the Lagrangian of this model, the axion couples directly only to the right-handed neutrinos. The couplings between the axion and the light neutrinos are induced through the diagonalization of the neutrino mass matrix in the type-I seesaw mechanism.
To obtain the standard axion--fermion couplings, we remove the axion field from the Yukawa terms in Eq.~\eqref{eq:Yukawa} by performing a field-dependent transformation
\begin{equation}
	\begin{pmatrix}
		f_1\\
		f_2\\
		f_3
	\end{pmatrix}
	\;\rightarrow\;
	\mathrm{e}^{-\mathrm{i}Ya/\lambda}
	\begin{pmatrix}
		f_1\\
		f_2\\
		f_3
	\end{pmatrix},
	\label{eq:f}
\end{equation}
where $f_i$ (with $i=1,2,3$) collectively denotes the three generations of fermions, specifically the $N_{iR}$, $l_{iR}$, and $L_i$ fields, and $a$ is the axion field, which is a linear combination of the phases of $\phi_{1,2}$,
\begin{equation}
	a=\frac{1}{\lambda}\!\left[\lambda_1(1+\chi)a_1+2\lambda_2\chi a_2\right],\qquad 
	\lambda=\sqrt{\lambda_1^2(1+\chi)^2+4\lambda_2^2\chi^2}.
\end{equation}
Here $Y$ is the PQ charge matrix in the three-generation space,
\begin{equation}
	Y=\begin{pmatrix}
		1&0&0\\
		0&\chi&0\\
		0&0&\chi
	\end{pmatrix}.\label{eq:Y}
\end{equation}
After the transformation in Eq.~\eqref{eq:f}, the axion--neutrino couplings in the mass basis take the form
\begin{equation}
	\mathcal{L}_{an}
	=\frac{\partial_\mu a}{2\lambda}\,
	\bar{n}\gamma^\mu\big(\mathrm{i}\operatorname{Im}\mathcal{Y}
	+\operatorname{Re}\mathcal{Y}\gamma_5\big)n,
	\label{eq:Lan1}
\end{equation}
where $n = (\nu, N)^T$ is the six-flavor Majorana neutrino multiplet in the mass basis, with $\nu$ and $N$ representing the light and heavy neutrino states, respectively. 
The matrix $\mathcal{Y}$ represents the PQ charge matrix of all the six-flavor neutrino fields in the mass basis, given by
\begin{equation}
	\mathcal{Y}= U_n^\dagger \begin{pmatrix}
		-Y&0\\
		0&Y
	\end{pmatrix} U_n.\label{eq:Un}
\end{equation}
Here, in the right-hand side of Eq.~\eqref{eq:Un}, each block of the middle matrices is a $3\times3$ matrix and $U_n$ is the $6\times6$ unitary matrix that diagonalizes the full neutrino mass matrix via
\begin{equation}
	U_n^TM_nU_n=\hat{M}_n,\label{eq:UnMn}
\end{equation}
with
\begin{equation}
	M_n=\begin{pmatrix}
		0&M^{(D)}\\
		{M^{(D)}}^T&M^{(R)}
	\end{pmatrix},\quad \hat{M}_n=
	\begin{pmatrix}
		\hat{M}_\nu&0\\
		0&\hat{M}_N
	\end{pmatrix},
\end{equation}
where $\hat{M}_\nu$ and $\hat{M}_N$ are the diagonal mass matrices for the light and heavy neutrinos, respectively.
	
For the light neutrino sector, the interactions in Eq.~\eqref{eq:Lan1} reduce to the following couplings:
\begin{equation}
	\mathcal{L}_{a\nu}
	=\frac{\partial_\mu a}{2\lambda}\,
	\bar{\nu}\gamma^\mu\big(\mathrm{i}\operatorname{Im}Y_\nu
	+\operatorname{Re}Y_\nu\gamma_5\big)\nu,\label{Lanu0}
\end{equation}
where $Y_\nu= U^\dagger Y U$ represents the PQ charge matrix of the light neutrinos in the mass basis.
	
Although the derivative interaction provides the most direct description of the axion--neutrino couplings, the equivalent (pseudo)scalar coupling form is also frequently used in the literature~\cite{Shin:1987xc,Pilaftsis_1994,Cely2017,Heeck2019}, and we will also employ this form later. Let us now derive this coupling form. Using integration by parts together with the equations of motion (EOMs), the derivative interaction can be rewritten as
\begin{equation}
	\mathcal{L}_{an}=\frac{\mathrm{i}a}{2\lambda}\,
	\bar{n}_i\!\left[(m_{n_i}+m_{n_j})\operatorname{Re}{\mathcal{Y}}_{ij}\gamma_5
	+\mathrm{i}(m_{n_i}-m_{n_j})\operatorname{Im}{\mathcal{Y}}_{ij}\right]\!n_j.
	\label{Lan2}
\end{equation}
Here, to distinguish the light- and heavy-neutrino masses, we denote the diagonal elements of $\hat{M}_\nu$ and $\hat{M}_N$ by $m_{\nu_i}$ and $m_{N_i}$, respectively, and collectively refer to them as the diagonal elements of $\hat{M}_n$, $m_{n_i}$, i.e.
\begin{equation}
m_{n_1},m_{n_2},m_{n_3}=m_{\nu_1},m_{\nu_2},m_{\nu_3},\quad 
m_{n_4},m_{n_5},m_{n_6}=m_{N_1},m_{N_2},m_{N_3}.
\end{equation}
In particular, in the light-neutrino sector the couplings reduce to
\begin{align}
	\mathcal{L}_{a\nu}
	&=-\frac{\mathrm{i}a}{2\lambda}\,
	\bar{\nu}_i\Big[(m_{\nu_i}+m_{\nu_j})\operatorname{Re}(Y_{\nu})_{ij}\gamma_5
	+\mathrm{i}(m_{\nu_j}-m_{\nu_i})\operatorname{Im}(Y_{\nu})_{ij}\Big]\nu_j
	\notag\\
	&=-\frac{\mathrm{i}a}{2\lambda}\big(\bar{\nu}_{L}O\nu_{R}-\bar{\nu}_{R}O^\dagger\nu_{L}\big),
	\label{Lanu2}
\end{align}
where
\begin{equation}
	O=2\hat{M}_\nu+(\chi-1)\big(\hat{M}_\nu\tilde{Y}_{\nu}^*+\tilde{Y}_{\nu} \hat{M}_\nu\big),\quad \tilde{Y}_{\nu}=
	U^\dagger 
	\begin{pmatrix}
		0&0&0\\
		0&1&0\\
		0&0&1
	\end{pmatrix}
	U.
\end{equation}
It's noteworthy that the off-diagonal terms in Eq.\eqref{Lanu2} are proportional to $(\chi-1)$.

Regarding the QCD axion sector, the most stringent astrophysical constraint is derived from the CMB free-streaming argument \cite{huang2018neutrinophilic,Hannestad2005}. This bound targets the neutrino transition process $\nu \to \nu' + a$, which constrains the off-diagonal couplings in Eq.~\eqref{Lanu2} as follows:
\begin{equation}
	\frac{m_{\nu_2}|(Y_{\nu})_{23}|}{2\lambda}, \, \frac{m_{\nu_1}|(Y_{\nu})_{13}|}{2\lambda}\lesssim10^{-11}.
\end{equation}
Taking a benchmark value of $\lambda \sim 10^9$ GeV, which aligns with the lower bound on the axion decay constant from astrophysical and cosmological considerations \cite{DILUZIO20201}, and assuming $(\chi - 1) \sim \mathcal{O}(1)$, the resulting couplings are approximately
\begin{equation}
	\frac{m_{\nu_2}|(Y_{\nu})_{23}|}{2\lambda}\sim\frac{m_{\nu_1}|(Y_{\nu})_{13}|}{2\lambda}\sim10^{-20}.
\end{equation}
These couplings are many orders of magnitude below the current experimental limits. Consequently, it implies that within the minimal realization of this model, the axion--neutrino interactions are significantly suppressed, remaining well beyond the sensitivity of foreseeable experimental probes.

A promising way to generate observable neutrino signatures is the clockwork mechanism, which can significantly enhance the axion--neutrino coupling by amplifying the PQ charge difference $(\chi-1)$. The technical implementation of this enhancement within our model is detailed in \ref{ap:clock}.

Beyond this enhancement, high-energy neutrinos from cosmic rays may provide another distinct signature through axion-modified neutrino oscillations.
The derivative axion--neutrino interactions in the flavor basis can be obtained directly from Eq.~\eqref{Lanu0} as
\begin{equation}
	\mathcal{L}_{a\nu}
	=-\frac{\partial_\mu a}{2\lambda}\,
	\bar{\nu}_{f}\gamma^\mu Y\gamma_5\nu_{f},
\end{equation}
where $\nu_f$ denotes the neutrino fields in the flavor basis.
Following the formalism presented in Ref.~\cite{huang2018neutrinophilic} for neutrino propagation in an axion background, we derive the corresponding effective Hamiltonian in the flavor basis:
\begin{equation}
	H=\frac{1}{2E}U\begin{pmatrix}
		0&0&0\\
		0&\Delta m^2_{21}&0\\
		0&0&\Delta m^2_{31}
	\end{pmatrix}U^\dagger+\frac{ \dot{a}(t)}{2\lambda}\begin{pmatrix}
		\chi-1&0&0\\
		0&0&0\\
		0&0&0
	\end{pmatrix}.\label{eq:Hofnu}
\end{equation}
where $E$ is the energy of the neutrino and we subtract the flavor universal part in the second term since it does not affect oscillations.
Here, the axion is treated as a classical field background
\begin{equation}
	a(t)=\frac{\sqrt{2\rho}}{m_a}\cos(m_a t),
\end{equation}
where $\rho$ is the cold axion energy density and $m_a$ is the axion mass.

Notably, the vacuum term in the effective Hamiltonian~\eqref{eq:Hofnu} is suppressed by the neutrino energy $1/E$, whereas the axion-induced potential remains independent of $E$. Consequently, the modifications introduced by the axion background become increasingly significant at high energies. If cold axion constitutes the dominant component of dark matter in the universe, then for $(\chi-1) \sim \mathcal{O}(1)$, the axion term becomes comparable to the vacuum term at a characteristic energy scale of
\begin{equation}
	E\sim\frac{\Delta m^2_{21}\lambda}{\sqrt{2\rho}}\sim10~\text{PeV},
\end{equation}
where we use $\sqrt{2\rho}\approx 2\times10^{-3}\text{eV}^2$ \cite{huang2018neutrinophilic,losada2023parametric} and $\lambda\sim10^9\text{GeV}$. This characteristic energy lies just close to the upper end of the energy range of cosmic neutrinos currently probed by the IceCube experiment.~\cite{Abbasi2025}.
In this regime, axion--neutrino interactions may substantially modify neutrino oscillation patterns, thereby altering the flavor composition of cosmic neutrino fluxes reaching Earth \cite{Song2021, Abbasi2025}.

However, since the axion-induced modification cannot be treated as a small perturbation in this regime, and the axion field is not approximately constant over the vast propagation distances of cosmic neutrinos, the resulting signatures are complex to evaluate. A precise numerical treatment of these effects is beyond the scope of the present study and will be addressed in future work.

Finally, it is noteworthy that a manifest difference arises if one employs the (pseudo)scalar couplings in Eq.~\eqref{Lanu2} to derive the effective Hamiltonian. Within the framework of the effective Dirac equation or the modified dispersion relation \cite{Ge2019, losada2023parametric, huang2018neutrinophilic}, the neutrino oscillation Hamiltonian is generally expressed as
\begin{equation}
	H\approx \frac{M_{\text{eff}}M_{\text{eff}}^\dagger}{2 E}+V.\label{eq:Happrox}
\end{equation}
The difference between these two types of interactions is apparent: the (pseudo)scalar couplings enter the effective mass matrix $M_{\text{eff}}$, resulting in an effect that is suppressed by the neutrino energy $1/E$ \cite{Ge2019, losada2023parametric, Dutta2024, Medhi2022}. In contrast, the derivative interaction in Eq.~\eqref{Lanu0} contribute to the matter potential $V$, which remains independent of the neutrino energy $E$ \cite{huang2018neutrinophilic}. 

From the perspective of quantum field theory, the derivative couplings in Eq.~\eqref{Lanu0} are formally equivalent to the (pseudo)scalar couplings in Eq.~\eqref{Lanu2} for on-shell particles \cite{Hannestad2005, Bauer2021, Bollig2020, Dias2014, Ernst2018}. However, within the approximate Hamiltonian treatment of Eq.~\eqref{eq:Happrox}, the functional differences in the resulting Hamiltonians may suggest a subtle non-equivalence when evaluating neutrino oscillations in a medium. 
Recent studies, such as Refs.~\cite{Lucente2021,Fiorillo2026,Carenza2021}, have explored this issue, demonstrating the equivalence of the two representations within the context of relativistic plasmas. Nevertheless, this equivalence is not general; e.g. it ceases to be valid when two Goldstone bosons are attached to one fermion line \cite{Choi1988}.

\subsection{The axion--charged lepton couplings}

\begin{figure}
	\centering
	\includegraphics[width=0.9\linewidth]{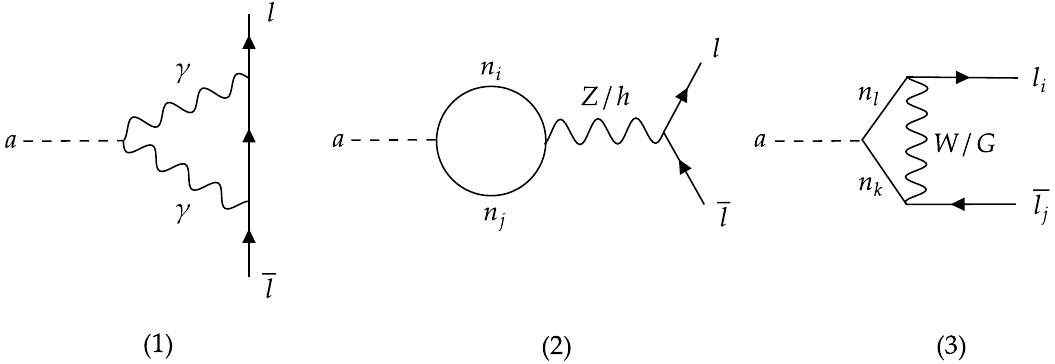}
	\caption{Feynman diagrams contributing to the axion--charged lepton couplings at one loop.\label{fig:Loops}}
\end{figure}

The coupling between the axions and charged leptons differs qualitatively from the axion--neutrino coupling, since no such interaction exists at tree level. This can be seen directly from the Yukawa terms in Eq.~\eqref{eq:Yukawa}. After performing the field redefinition in Eq.~\eqref{eq:f}, one might naively expect a tree-level interaction of the form
\begin{equation}
	\mathcal{L}_{al}=\frac{\partial_\mu a}{\lambda}\bar{l}\gamma^\mu Yl,
	\label{eq:Lal}
\end{equation}
to be present. However, by integrating by parts and using the EOMs, one finds that this term vanishes identically. Therefore, there is indeed no tree-level axion--charged lepton coupling in this model. Nevertheless, such a coupling is generated at the one-loop level.
The axion--charged lepton couplings at one loop arise from the Feynman diagrams shown in Fig.~\ref{fig:Loops}~\cite{Shin:1987xc,Pilaftsis_1994,Cely2017,Heeck2019}.

\subsubsection{The expression of the amplitudes}
As shown in Ref.~\cite{Giannotti2017}, the neutrino-mediated contributions dominate over the photon-mediated ones. These neutrino-mediated couplings were previously computed in Refs.~\cite{Shin:1987xc,Cely2017,Heeck2019} within flavor-universal models under the assumption that there is no intermixing angles in heavy neutrino sector, i.e.\ that $M^{(R)}$ is diagonal. However, this assumption cannot be realized in the present model, since the condition $M^{(R)}_{11}=0$ is required.
We therefore follow the general approach of Ref.~\cite{Pilaftsis_1994}, which does not rely on vanishing heavy-neutrino mixing angles, to compute the axion couplings to charged leptons in our framework.

\begin{figure}
	\centering
	\includegraphics[width=0.9\linewidth]{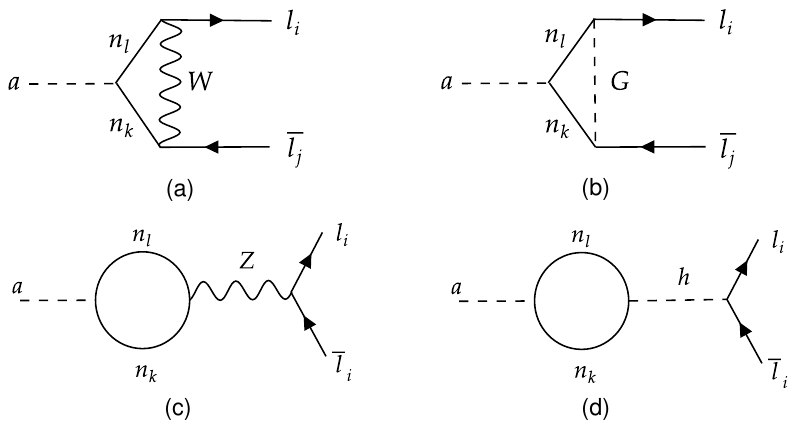}
	\caption{The four Feynman diagrams evaluated (in the Feynman gauge) in this work.\label{fig:Loops2}}
\end{figure}

In what follows, we evaluate the Feynman diagrams shown in Fig.~\ref{fig:Loops2} in the Feynman gauge. 
Diagram (d) of Fig.~\ref{fig:Loops2} actually represents the sum of three diagrams obtained by replacing $h$ with $h'$, $\rho_1'$ and $\rho_2'$, respectively. Here, $h'$, $\rho_1'$ and $\rho_2'$ are the mass eigenstates resulting from the mixing among $h$, $\rho_1$ and $\rho_2$, induced by scalar-potential terms such as
\begin{equation}
	|H|^2|\phi_1|^2,\quad |H|^2|\phi_2|^2.
\end{equation}
Accordingly, the interaction eigenstate $h$ can be expanded in terms of the mass eigenstates as
\begin{equation}
 h=\alpha_1\rho_1'+\alpha_2\rho_2'+\alpha_3 h',\quad \alpha_1^2+\alpha_2^2+\alpha_3^2=1.
\end{equation}
However, as will be shown below, the amplitude corresponding to diagram (d) vanishes for each mass eigenstate separately. Therefore, the mixing among $h$, $\rho_1$ and $\rho_2$ does not affect the final result. For clarity, we first neglect this mixing by setting
\begin{equation}
	\alpha_1=\alpha_2=0,\quad \alpha_3=1.
\end{equation}
At the end of Sec.~\ref{sec:Md}, we briefly explain why including the scalar mixing does not modify the final result.

Before presenting the calculation, we emphasize that, unless explicitly stated otherwise, we keep the discussion as model-independent as possible. The PQ charge matrix $Y$, the mass matrices $M^{(D)}$ and $M^{(R)}$, and the mixing matrices $U_n$ and $U_{lL}$ are therefore not specified explicitly. Instead, besides the fundamental framework of the type-I seesaw and the KSVZ model, we only assume that they satisfy the following conditions:
\begin{subequations}\label{eq:assumption}
	\begin{align}
		Y&\ \text{is diagonal}, \label{eq:assumptiona}\\
		[Y,M^{(D)}]&=[Y,U_{lL}]=0. \label{eq:assumptionb}
	\end{align}
\end{subequations}
These conditions are satisfied by a broad class of $\nu$KSVZ models.

The interactions beyond the SM appearing in Fig.~\ref{fig:Loops2}, in addition to the axion--neutrino coupling given in Eq.~\eqref{Lan2} \footnote{Notably, the derivation of Eq.~\eqref{Lan2} relies only on the assumption \eqref{eq:assumptiona}.}, are
\begin{equation}
	\mathcal{L}_{Wnl}=\frac{e}{\sqrt{2}\sin\theta_w}W^-_\mu \bar{l}_{i}\gamma^\mu B_{ij}P_Ln_{j}+h.c.,
	\label{LWnl}
\end{equation}
\begin{equation}
	\mathcal{L}_{Gnl}=\frac{\sqrt{2}}{\lambda_H}G^-\bar{l}_i\big(m_{l_i}B_{ij}P_L-B_{ij}m_{n_j}P_R\big)n_j+h.c.,
	\label{LGnl}
\end{equation}
\begin{equation}
	\mathcal{L}_{Zn}=\frac{e}{4\cos\theta_w\sin\theta_w}Z_\mu\bar{n}_{i}\gamma^\mu \big(\mathrm{i}\operatorname{Im}C_{ij}-\gamma_5\operatorname{Re}C_{ij}\big)n_{j},
	\label{LZn}
\end{equation}
\begin{equation}
	\mathcal{L}_{hn}=\frac{h}{2\lambda_H}\bar{n}_i \Big[(m_{n_i}+m_{n_j})\operatorname{Re}C_{ij}
	+\mathrm{i}\gamma_5(m_{n_j}-m_{n_i})\operatorname{Im}C_{ij}\Big]n_j,
	\label{Lhn}
\end{equation}	
where $e$ is the electromagnetic coupling strength, $\theta_w$ is the Weinberg angle, $m_l$ and $m_n$ denote the charged-lepton and neutrino masses respectively, $P_L=\tfrac12(1-\gamma_5)$, and $P_R=\tfrac12(1+\gamma_5)$. 
The matrices $B_{ij}$ and $C_{ij}$ are defined as
\begin{equation}
	B_{ij}=\sum_{k=1}^3 (U_{lL}^*)_{ki}(U_n^*)_{kj},
\end{equation}
\begin{equation}
	C_{ij}=\sum_{k=1}^3 (U_n)_{ki}(U_n^*)_{kj}.
\end{equation}
Using the Lagrangians in Eqs.~\eqref{Lan2} and \eqref{LWnl}--\eqref{Lhn}, one can compute the individual amplitudes corresponding to the diagrams shown in Fig.~\ref{fig:Loops2}. 
In what follows, we neglect terms proportional to the small ratio $m_l^2/m_W^2$, where $m_W$ denotes the mass of the $W$ boson. The calculation results are as follows:

\begin{align}
	M_{a}=&\frac{g_1^2}{32\pi^2\lambda}\bar{v}^i_s(p_1)\Big\{\big[m_{l_j}(1+\gamma_5)-m_{l_i}(1-\gamma_5)\big]\omega^{S}_{ij}
	+\big[m_{l_j}(1+\gamma_5)+m_{l_i}(1-\gamma_5)\big]\omega_{ij}^{A}\Big\}u^r_{j}(p_3),\label{eq:Ma}\\
	M_{b}=&-\frac{1}{16\pi^2\lambda_H^2\lambda}\bar{v}^i_s(p_1)\Big\{\big[m_{l_j}(1+\gamma_5)-m_{l_i}(1-\gamma_5)\big]\eta^{S}_{ij}
	+\big[m_{l_j}(1+\gamma_5)+m_{l_i}(1-\gamma_5)\big]\eta_{ij}^{A}\Big\}u^r_{j}(p_3),\label{eq:Mb}\\
	M_{c}=&\frac{m_{l_i}\bar{v}^{s}_i(p_1)\gamma_5u^{r}_i(p_3)}{(4\pi)^2\lambda_H^2\lambda}F_c,\label{eq:Mc}\\
	M_d=&\frac{m_{l_i}\bar{v}_i^s (p_1)u_i^r(p_3)}{(4\pi)^2m_h^2\lambda^2_H\lambda}F_d,\label{eq:Md}
\end{align}
where 
\begin{align}
	\omega^S_{ij}&=\frac{1}{2}B_{il}B^*_{jk}\Big\{\big[(\lambda_l+\lambda_k)\mathcal{Y}^*_{lk}+2\sqrt{\lambda_l\lambda_k}\mathcal{Y}_{lk}\big]I_1(\lambda_l,\lambda_k)+2\mathcal{Y}^*_{lk}h(\lambda_l,\lambda_k)\Big\},\label{eq:omegaS}\\
	\omega^A_{ij}&=\frac{1}{2}B_{il}B^*_{jk}\Big\{(\lambda_l-\lambda_k)\mathcal{Y}^*_{lk}I_1(\lambda_l,\lambda_k)+2\mathcal{Y}^*_{lk}p(\lambda_l,\lambda_k)\Big\},\label{eq:omegaA}\\
	\eta^S_{ij}&= B_{il}B^*_{jk}\Big\{m_{n_l}m_{n_k}\mathcal{Y}_{lk}h(\lambda_l,\lambda_k)+2C_{UV}\big[m_{n_l}m_{n_k}\mathcal{Y}_{lk}+\frac{1}{2}(m_{n_l}^2+m_{n_k}^2)\mathcal{Y}_{lk}^*\big]\Big\},\label{eq:etaS}\\
	\eta^A_{ij}&= B_{il}B^*_{jk}\Big[m_{n_l}m_{n_k}\mathcal{Y}_{lk}p(\lambda_l,\lambda_k)+C_{UV}(m_{n_l}^2-m_{n_k}^2)\mathcal{Y}_{lk}^*\Big],\label{eq:etaA}\\
	F_c&=-\Big\{\big(m_{n_l}^2-m_{n_k}^2\big)\big(C_{lk}\mathcal{Y}_{lk}+C^*_{lk}\mathcal{Y}^*_{lk}\big)\Big[\frac{1}{2}C'_{UV}-L_1(\lambda_l,\lambda_k)\Big]\notag\\
	&+m_{n_k}\big[m_{n_l}\big(C_{lk}\mathcal{Y}^*_{lk}+C^*_{lk}\mathcal{Y}_{lk}\big)+m_{n_k}\big(C_{lk}\mathcal{Y}_{lk}+C^*_{lk}\mathcal{Y}^*_{lk}\big)\big]\big[C'_{UV}-L_0(\lambda_l,\lambda_k)\big]\Big\},\label{eq:Fc}\\
	F_d&=(m_{n_l}^2-m_{n_k}^2)\Big\{-m_{n_l}m_{n_k}\big(C_{lk}\mathcal{Y}_{lk}^*-C^*_{lk}\mathcal{Y}_{lk}\big)\big[C'_{UV}-L_0(\lambda_l,\lambda_k)\big]\notag\\
	&+\big(C_{lk}\mathcal{Y}_{lk}-C^*_{lk}\mathcal{Y}^*_{lk}\big)\Big[\big(m_{n_l}^2+m_{n_k}^2\big)\Big(C'_{UV}+\frac{1}{2}\Big)-2\big(m_{n_l}^2-m_{n_k}^2\big)L_1(\lambda_l,\lambda_k)
	-2m_{n_k}^2L_0(\lambda_l,\lambda_k)\Big]\Big\}\label{eq:Fd},
\end{align}
and
\begin{align}
	\lambda_{l,k}&=\frac{m_{n_{l,k}}^2}{m_W^2},\\
	h(\lambda_l,\lambda_k)&=\frac{1}{2}(\lambda_{k}-\lambda_l)[I_3(\lambda_l,\lambda_k)-I_2(\lambda_l,\lambda_k)],\\
	p(\lambda_l,\lambda_k)&=\frac{1}{2}(\lambda_{k}-\lambda_l)[I_3(\lambda_l,\lambda_k)+I_2(\lambda_l,\lambda_k)],\\
	C'_{UV}&= \Big(\frac{1}{\epsilon}-\gamma_E+\ln4\pi-\ln\frac{m_W^2}{\mu^2}\Big),\\
	C_{UV}&=-\frac{1}{2}\Big(C'_{UV}-\frac{1}{2}\Big)+C_{\text{fin}}(\lambda_l,\lambda_k).
\end{align}
Here $ g_1 $ is the gauge coupling of $ SU(2)_L $, and the indices $l,k$ run from 1 to 6. The analytical expressions of the one-loop functions $I_1$, $I_2$, $I_3$, $L_0$, $L_1$, and $C_{\text{fin}}$ are given in \ref{ap:calc}. 

\subsubsection{Ultraviolet finiteness}
The UV divergent parts of the amplitudes in Eqs.~\eqref{eq:Mb}--\eqref{eq:Md}, namely the terms proportional to $C'_{UV}$ in the coefficients \eqref{eq:etaS}--\eqref{eq:Fd}, cancel as a consequence of the following matrix identities:
\begin{align}
	&B\mathcal{Y}^*=-YB,\label{eq:id1}\\
	&B\hat{M}_n\mathcal{Y}=YB\hat{M}_n,\label{eq:id2}\\
	&C\mathcal{Y}^*=\mathcal{Y}^*C,\label{eq:id3}\\
	&C\hat{M}_n\mathcal{Y}=-C\mathcal{Y}^*\hat{M}_n.\label{eq:id4}
\end{align}
The proof of these identities is presented in \ref{ap:id}.

To demonstrate this cancellation explicitly, we first extract the UV-divergent parts of the coefficients \eqref{eq:etaS}--\eqref{eq:Fd}. Since $C'_{UV}$ is independent of the summation indices $l$ and $k$, these UV-divergent terms can be written in matrix form:
\begin{align}
	\eta^S_{ij}\supset&-\Big(C'_{UV}-\frac{1}{2}\Big)\Big[ B\hat{M}_n\mathcal{Y}\hat{M}_nB^\dagger+\frac{1}{2}\Big(B\hat{M}_n^2\mathcal{Y}^*B^\dagger+B\mathcal{Y}^*\hat{M}_n^2B^\dagger\Big)\Big]_{ij},\\
	\eta^A_{ij}\supset& -\frac{1}{2}\Big(C'_{UV}-\frac{1}{2}\Big)\big(B\hat{M}_n^2\mathcal{Y}^*B^\dagger-B\mathcal{Y}^*\hat{M}_n^2B^\dagger\big)_{ij}\\
	F_c\supset &-\frac{1}{2}C'_{UV}\big[\operatorname{Tr}(\hat{M}_nC\mathcal{Y}^*\hat{M}_n)-\operatorname{Tr}(C\hat{M}_n^2\mathcal{Y}^*)\big]
	-C'_{UV}\big[\operatorname{Tr}(\hat{M}_nC\hat{M}_n\mathcal{Y})+\operatorname{Tr}(C\hat{M}_n^2\mathcal{Y}^*)\big]+c.c.,\\
	F_d\supset&-C'_{UV}\big[\operatorname{Tr}(\hat{M}_n^3C\hat{M}_n\mathcal{Y})-\operatorname{Tr}(\hat{M}_nC\hat{M}_n^3\mathcal{Y})\big]
	+\Big(C'_{UV}+\frac{1}{2}\Big)\big[\operatorname{Tr}(\hat{M}_n^4C\mathcal{Y}^*)-\operatorname{Tr}(C\hat{M}_n^4\mathcal{Y}^*)\big]-c.c..
\end{align}
It is then straightforward to verify that these divergent terms vanish because the identities below are satisfied	
\begin{equation}
	-B\hat{M}_n\mathcal{Y}\hat{M}_nB^\dagger=B\hat{M}_n^2\mathcal{Y}^*B^\dagger=B\mathcal{Y}^*\hat{M}_n^2B^\dagger,\label{eq:idB}
\end{equation}
\begin{equation}
	\operatorname{Tr}\big(\hat{M}_nC\mathcal{Y}^*\hat{M}_n\big)=\operatorname{Tr}\big(C\hat{M}^2_n\mathcal{Y}^*\big)=-\operatorname{Tr}\big(\hat{M}_nC\hat{M}_n\mathcal{Y}\big),\label{eq:idTr1}
\end{equation}
\begin{equation}
	\operatorname{Tr}\big(\hat{M}_n^3C\hat{M}_n\mathcal{Y}\big)=\operatorname{Tr}\big(\hat{M}_nC\hat{M}^3_n\mathcal{Y}\big)=	-\operatorname{Tr}\big(\hat{M}_n^4C\mathcal{Y}^*\big)=-\operatorname{Tr}\big(C\hat{M}^4_n\mathcal{Y}^*\big).\label{eq:idTr2}
\end{equation}
The identity \eqref{eq:idB} follows directly from the identities \eqref{eq:id1} and \eqref{eq:id2}, together with
\begin{equation}
	[B\hat{M}_n^2B^\dagger,Y]=0,
\end{equation}
this commutator vanishes because
\begin{equation}
	B\hat{M}_n^2B^\dagger=U_{lL}^\dagger M^{(D)}{M^{(D)}}^\dagger U_{lL},
\end{equation}
and the assumption \eqref{eq:assumptionb}. The identities \eqref{eq:idTr1} and \eqref{eq:idTr2} follow directly from the identities \eqref{eq:id3} and \eqref{eq:id4}.

\subsubsection{Vanishing of $M_d$ irrespective of scalar mixing}\label{sec:Md}
Here, we show that the coefficient $F_d$, and hence the amplitude $M_d$, vanishes identically. After the cancellation of the UV-divergent contributions, $F_d$ can be written as
\begin{align}
	F_d&=-(m_{n_l}^2-m_{n_k}^2)\big\{\big[T_0^{(1)}(l,k)+T_0^{(2)}(l,k)\big]L_0(\lambda_l,\lambda_k)
	+T_1(l,k)L_1(\lambda_l,\lambda_k)\big\},
\end{align}
where
\begin{align}
	T_0^{(1)}(l,k)&=-m_{n_l}m_{n_k}\big(C_{lk}\mathcal{Y}_{lk}^*-C^*_{lk}\mathcal{Y}_{lk}\big),\\
	T_0^{(2)}(l,k)&=2m_{n_k}^2\big(C_{lk}\mathcal{Y}_{lk}-C^*_{lk}\mathcal{Y}^*_{lk}\big),\\
	T_1(l,k)&=2\big(m_{n_l}^2-m_{n_k}^2\big)\big(C_{lk}\mathcal{Y}_{lk}-C^*_{lk}\mathcal{Y}^*_{lk}\big).
\end{align}
Using the matrix identities \eqref{eq:id3} and \eqref{eq:id4}, one finds that, after summation over $l$ and $k$,
\begin{align}
	(m_{n_l}^2-m_{n_k}^2)T_0^{(1)}(l,k)L_0(\lambda_l,\lambda_k)&=0,\\
	(m_{n_l}^2-m_{n_k}^2)\big[T_0^{(2)}(l,k)L_0(\lambda_l,\lambda_k)+T_1(l,k)L_1(\lambda_l,\lambda_k)\big]&=0.
\end{align}
Therefore, $F_d$ vanishes identically.
	
In deriving this result, we have neglected the mixing among $h$, $\rho_1$, and $\rho_2$. Thus, the vanishing of $F_d$ first implies that the amplitude corresponding to diagram (d) in Fig.~\ref{fig:Loops2} vanishes in the absence of scalar mixing.

When scalar mixing is included, the amplitude becomes
\begin{equation}
	M_d=\frac{m_{l_i}\bar{v}_i^s (p_1)u_i^r(p_3)}{(4\pi)^2\lambda^2_H\lambda}\Big(\frac{\alpha_1^2}{m_{\rho_1'}^2}+\frac{\alpha_2^2}{m_{\rho_2'}^2}+\frac{\alpha_3^2}{m_{h'}^2}\Big)F_d.
\end{equation}
Since each mass-eigenstate contribution is proportional to the same vanishing coefficient $F_d$, the amplitude $M_d$ vanishes for an arbitrary scalar mixing pattern.

\subsubsection{Effective on-shell couplings}\label{sec:Lal}
We next evaluate the leading-order contributions to the remaining nonvanishing coefficients in Eqs.~\eqref{eq:omegaS}--\eqref{eq:Fc}.
	
The crucial step is to decompose the sums over the internal-neutrino indices $l$ and $k$ into contributions from the light- and heavy-neutrino sectors. We separate the ranges of these indices as
\begin{equation}
	l,k=1,2,3 \quad\text{and} \quad l,k=4,5,6.
\end{equation}
The first range corresponds to the light-neutrino sector, whose indices will be denoted by $\nu$ and $\nu'$, while the second corresponds to the heavy-neutrino sector, whose indices will be denoted by $N$ and $N'$.

Since each coefficient in Eqs.~\eqref{eq:omegaS}--\eqref{eq:Fc} involves a double sum over $l$ and $k$, it can be decomposed into light--light, light--heavy, heavy--light, and heavy--heavy contributions. For example,
\begin{align}
	\omega^S_{ij}&=\sum_{l,k}\omega^S_{ij}(l,k)\nonumber\\
	&=\sum_{\nu,\nu'}\omega^S_{ij}(\nu,\nu')+\sum_{\nu,N}[\omega^S_{ij}(\nu,N)+\omega^S_{ij}(N,\nu)]+\sum_{N,N'}\omega^S_{ij}(N,N').
\end{align}
The light--light contribution is always suppressed by the squared light-neutrino masses. Therefore, only the light--heavy, heavy--light, and heavy--heavy contributions enter at leading order.
	
We first evaluate the leading-order contributions to $\omega^S_{ij}$ and $\omega^A_{ij}$. The neutrino mixing matrix $U_n$ can be expanded in powers of
\begin{equation}
	\epsilon=M^{(D)}{M^{(R)}}^{-1},
\end{equation}
as
\begin{equation}
	U_n=\begin{pmatrix}
		U_\nu^*&\epsilon^*U_N\\
		-\epsilon^TU_\nu^*&U_N
	\end{pmatrix}+\mathcal{O}(\epsilon^2),
\end{equation}
where the unitary matrix $U_N$ is defined through
\begin{equation}
	\hat{M}_N=U_N^T M^{(R)} U_N.\label{eq:MNUN}
\end{equation}
It then follows that
\begin{equation}
	B_{i\nu}B^*_{jN}\mathcal{Y}^{(*)}_{\nu N}\sim B_{iN}B^*_{j\nu}\mathcal{Y}^{(*)}_{ N\nu}\sim B_{iN}B^*_{jN'}\mathcal{Y}^{(*)}_{NN'}\sim \epsilon^2.
\end{equation}
Moreover, the leading behavior of the loop integrals $I_1$, $I_2$, and $I_3$ is
\begin{equation}
	I_{1,2,3}(\nu,N)\sim I_{1,2,3}(N,\nu)\sim I_{1,2,3}(N,N')\sim \frac{m_W^2}{m_{n_N}^2}\sim \frac{m_W^2}{m^2_{n_{N'}}}.
\end{equation}
These loop integrals compensate the explicit heavy-neutrino mass factors appearing in $\omega^S_{ij}$ and $\omega^A_{ij}$. Consequently,
\begin{equation}
	\omega^S_{ij}\sim \omega^A_{ij}\sim \epsilon^2.
\end{equation}
Thus, both $\omega^S_{ij}$ and $\omega^A_{ij}$ are suppressed by two powers of the small seesaw expansion parameter $\epsilon$.
	
By contrast, the suppression by $\epsilon^2$ is compensated in $\eta^S_{ij}$ and $\eta^A_{ij}$ by the two additional powers of heavy-neutrino masses appearing in these coefficients. The leading contributions to $\eta^S_{ij}$ from the three relevant sectors are
\begin{align}
	\sum_{\nu,N}\eta^S_{ij}(\nu,N)&=\frac{1}{4}(3-2\ln\Lambda_l)G_{jl}^*\Big[Y_{i}G_{il}+G_{ik}(Y_N)_{lk}\frac{m_{N_{l}}}{m_{N_{k}}}\Big]+\mathcal{O}(\epsilon^2),\label{eq:nuN}\\
	\sum_{\nu,N}\eta^S_{ij}(N,\nu)&=\frac{1}{4}(3-2\ln\Lambda_l)G_{il}\Big[Y_jG_{jl}^*+G^*_{jk}(Y_N)^*_{lk}\frac{m_{N_l}}{m_{N_{k}}}\Big]+\mathcal{O}(\epsilon^2),\label{eq:Nnu}\\
	\sum_{N,N'}\eta^S_{ij}(N,N')&= G_{il}G^*_{jk}\Big\{(Y_N)_{lk}h(\Lambda_l,\Lambda_{k})+2C_{\text{fin}}(\Lambda_l,\Lambda_k)
	\Big[(Y_N)_{lk}+\frac{1}{2}\Big(\frac{m_{N_l}}{m_{N_{k}}}+\frac{m_{N_{k}}}{m_{N_l}}\Big)(Y_N)_{lk}^*\Big]\Big\},\label{eq:NN'}
\end{align}
where 
\begin{equation}
	G=U^\dagger_{lL}M^{(D)}U_N,\quad Y_N=U_N^\dagger YU_N,\quad\Lambda_{l,k}=\frac{m_{N_{l,k}}^2}{m_W^2}.
\end{equation}
For notational convenience, in Eqs.~\eqref{eq:nuN}--\eqref{eq:NN'}, the indices $\nu=1,2,3$ and $N,N'=4,5,6$ are each relabeled as $l,k=1,2,3$, allowing the relevant parts of the $6\times6$ matrix $\mathcal{Y}$ and the $3\times6$ matrix $B$ to be expressed in terms of the $3\times3$ matrices $G$, $Y$ and $Y_N$.
	
Adding the three contributions, we obtain the leading-order expression for $\eta^S_{ij}$:
\begin{align}
	\eta^S_{ij}=&\frac{1}{4}(3-2\ln\Lambda_l)G_{il}G_{jl}^*(Y_i+Y_j)\nonumber\\
	&+G_{il}G_{jk}^*(Y_N)_{lk}\Big[-1+\frac{(3\Lambda_l-\Lambda_k)\ln\Lambda_l-(3\Lambda_k-\Lambda_l)\ln\Lambda_k}{4(\Lambda_l-\Lambda_k)}\Big]\nonumber\\
	&+G_{il}G_{jk}^*(Y_N)_{lk}^*\sqrt{\Lambda_l\Lambda_k}\frac{\ln\Lambda_l-\ln\Lambda_k}{\Lambda_l-\Lambda_k}+\mathcal{O}(\epsilon^2).\label{eq:LOetaS}
\end{align}
An analogous calculation gives the leading-order expression for $\eta^A_{ij}$:
\begin{equation}
	\eta^A_{ij}=\frac{3}{2}(U_{lL}^\dagger M^{(D)}[Y,R]{M^{(D)}}^\dagger U_{lL})_{ij}+\mathcal{O}(\epsilon^2),\label{eq:LOetaA}
\end{equation}
where
\begin{equation}
	R=U_N\ln\frac{\hat{M}_N}{m_W}U_N^\dagger.
\end{equation}
	
Finally, the coefficient $F_c$ can be simplified to
\begin{align}
	F_c=&F_1(m_{n_l},m_{n_k}) \big(C_{lk}\mathcal{Y}_{mlk}+C^*_{lk}\mathcal{Y}^*_{mlk}\big)
	+F_2(m_{n_l},m_{n_k})\big(C_{lk}\mathcal{Y}^*_{mlk}+C^*_{lk}\mathcal{Y}_{mlk}\big),
\end{align}
where
\begin{align}
	F_1(m_{n_l},m_{n_k})=&-\frac{1}{4}(m_{n_l}^2+m_{n_k}^2)+\frac{m_{n_l}^4\ln \lambda_l-m_{n_k}^4\ln \lambda_k}{2(m_{n_l}^2-m_{n_k}^2)},\\
	F_2(m_{n_l},m_{n_k})=&-m_{n_l}m_{n_k}+\frac{m_{n_l}m_{n_k}}{m_{n_l}^2-m_{n_k}^2}(m_{n_l}^2\ln \lambda_l-m_{n_k}^2\ln \lambda_k).
\end{align}
One can verify that both $F_1$ and $F_2$ are symmetric under the exchange of their two arguments, and that
\begin{align}
	&F_1(m_{n_\nu},m_{n_N})\sim F_1(m_{n_N},m_{n_{N'}})\sim F_2(m_{n_N},m_{n_{N'}})\sim m_{n_N}^2,\\
	&C_{\nu N}\mathcal{Y}_{\nu N}\sim C_{N\nu}\mathcal{Y}_{N\nu}\sim C_{NN'}\mathcal{Y}_{NN'}\sim \epsilon^2
\end{align}
Therefore, the suppression by $\epsilon^2$ is again compensated by the heavy-neutrino mass factors, and the leading-order contribution to $F_c$ is
\begin{align}
F_c=&(1-2\ln\Lambda_l)m_{N_l}^2\operatorname{Re}A_{ll}\nonumber\\
&+2F_1(m_{N_l},m_{N_k})\operatorname{Re}\big[D_{lk}(Y_N)_{lk}\big]
+2F_2(m_{N_l},m_{N_k})\operatorname{Re}\big[D_{lk}(Y_N)^*_{lk}\big]+\mathcal{O}(\epsilon^2),\label{eq:LOFc}
\end{align}
where
\begin{equation}
	A=U_N^T(Y\epsilon^\dagger\epsilon+\epsilon^\dagger Y\epsilon) U^*_N,\quad D=U_N^T\epsilon^\dagger\epsilon U_N^*.
\end{equation}
For notational convenience, the indices $\nu=1,2,3$ and $N,N'=4,5,6$ are again relabeled as $l,k=1,2,3$, allowing the relevant parts of the $6\times6$ matrices $\mathcal Y$ and $C$ to be expressed in terms of the $3\times3$ matrices $A$, $D$ and $Y_N$.
	
Finally, the leading on-shell axion--charged lepton coupling, which receives contributions only from $M_b$ and $M_c$, is given by
\begin{align}
	\mathcal{L}_{al}=\frac{\mathrm{i}}{16\pi^2\lambda_H^2\lambda}a\bar{l}_i\Big\{&\big[m_{l_j}(1+\gamma_5)-m_{l_i}(1-\gamma_5)\big]\Big(\eta^{S}_{ij}-\frac{1}{2}\delta_{ij}F_c\Big)\nonumber\\
	&+\big[m_{l_j}(1+\gamma_5)+m_{l_i}(1-\gamma_5)\big]\eta_{ij}^{A}\Big\}l_j,
\end{align}\label{eq:Lal1}
where the coefficients $\eta^{S}_{ij}$, $\eta^{A}_{ij}$ and $F_c$ are given by Eq.\eqref{eq:LOetaS}, Eq.\eqref{eq:LOetaA} and Eq.\eqref{eq:LOFc}, respectively. We note that the above one-loop result is obtained in the Feynman gauge, an explicit demonstration of gauge independence is not included in this work. Nevertheless, the result is largely model independent and applies to a broad class of $\nu$KSVZ models, provided that the assumptions in Eq.~\eqref{eq:assumption} are satisfied.

As a consistency check, in the flavor-universal limit this result reduces to the well-known expression obtained in Refs.~\cite{Cely2017,Heeck2019}:
\begin{equation}
	\mathcal{L}_{al}
	=\frac{\mathrm{i}a}{16\pi^2\lambda_H^2\lambda}\,
	\bar{l}_i\!\left[(m_{l_j}-m_{l_i})K_{ij}
	+\gamma_5(m_{l_j}+m_{l_i})\!\left(K_{ij}-\tfrac{1}{2}\delta_{ij}\operatorname{Tr}K\right)\right]\!l_j,
	\label{eq:Lal2}
\end{equation}	
where
\begin{equation}
	K=M^{(D)}{M^{(D)}}^\dagger.
\end{equation}
This reduction can be verified by taking the limit $U_{lL}=U_N=2Y=1$ and neglecting the $\mathcal{O}(\epsilon^2)$ terms, yielding
\begin{equation}
	\eta^S_{ij}=K_{ij},\quad \eta^A_{ij}=0,\quad F_c=\operatorname{Tr}K.
\end{equation}

\subsubsection{Phenomenological consequences and constraints}
The most significant difference between the effective Lagrangian in Eq.~\eqref{eq:Lal1} and that of the flavor-universal models~\cite{Pilaftsis_1994,Cely2017,Heeck2019} is the presence of the nonvanishing coefficient $\eta^A_{ij}$.
A nonvanishing $\eta^A_{ii}$ would induce a CP-violating diagonal axion--charged lepton coupling, which is subject to much stronger experimental constraints than its CP-conserving counterpart~\cite{Hare2020}. Under the assumptions in Eq.~\eqref{eq:assumption} and neglecting the higher-order $\mathcal{O}(\epsilon^2)$ terms, one can show that $\eta^A_{ii}$ always vanishes.  
	
To this end, we first define
\begin{equation}
	\Delta\equiv[Y,R],\quad X\equiv U_{lL}^\dagger M^{(D)}.
\end{equation}
The matrix elements of $\Delta$ can be expressed as
\begin{equation}
	\Delta_{ij}=(Y_i-Y_j)R_{ij}.
\end{equation}
Moreover, since $X$ commutes with $Y$,
\begin{equation}
	X_{ij}=0\quad \text{or}\quad  Y_i=Y_j.
\end{equation}
Therefore, after neglecting the higher-order $\mathcal{O}(\epsilon^2)$ terms, $\eta^A_{ii}$ can be expressed as
\begin{equation}
	\eta^A_{ii}=\frac{3}{2}(X\Delta X^\dagger)_{ii}=\frac{3}{2}(Y_j-Y_k)R_{jk}X_{ij}X^*_{ik}.
\end{equation}
A nonvanishing contribution to $\eta^A_{ii}$ requires
\begin{equation}
	X_{ij},X_{ik}\neq0.
\end{equation}
Since $X_{ij}\neq0$ implies $Y_i=Y_j$, and $X_{ik}\neq0$ implies $Y_i=Y_k$, one immediately obtains
\begin{equation}
	Y_j-Y_k=0.
\end{equation}
Therefore, neglecting the $\mathcal{O}(\epsilon^2)$ terms,
\begin{equation}
	\eta^A_{ii}=0.
\end{equation}
Consequently, although the flavored model predicts additional off-diagonal couplings, the diagonal CP-violating axion--charged lepton interactions vanish up to the order considered, and possible nonzero contributions only arise from $\mathcal{O}(\epsilon^2)$ corrections.

To compare our results with existing experimental constraints, we rewrite the effective Lagrangian~\eqref{eq:Lal1} in the standard form adopted in Ref.~\cite{Bauer_2022},
\begin{align}
	\mathcal{L}_{al}&=-\frac{\mathrm{i}a}{2f_a}\big[(m_{l_i}-m_{l_j})(k_f+k_F)_{ij}\bar{l}_il_j
	+(m_{l_i}+m_{l_j})(k_f-k_F)_{ij}\bar{l}_i\gamma_5l_j\big],\label{eq:Lal3}
\end{align}
where
\begin{align}
	(k_f)_{ij}&=\frac{m_{l_j}m_{l_i}}{m_{l_j}^2-m_{l_i}^2}\frac{\eta^A_{ij}}{4\pi^2N\lambda_H^2},\\
	(k_F)_{ij}&=\frac{1}{8\pi^2N\lambda_H^2}\Big[\eta^S_{ij}+\frac{m^2_{l_j}+m^2_{l_i}}{m^2_{l_j}-m^2_{l_i}}\eta^A_{ij}\Big],\\
	(k_f)_{ii}-(k_F)_{ii}&=-\frac{1}{8\pi^2N\lambda_H^2}\Big(\eta^S_{ii}-\frac{1}{2}F\Big),
\end{align}
and $i\neq j$ is required. And the number $N$ is defined by 
\begin{equation}
	N=\frac{\lambda}{f_a}.
\end{equation}
With the choice of the anomaly term in Eq.~\eqref{eq:anomaly}, $N=\chi_{\phi_1}$.

From Eq.~\eqref{eq:Lal3}, the effective axion couplings are conventionally defined as~\cite{Giannotti2017,DILUZIO20201,Bauer_2022,Hare2020}
\begin{align}
	g_{al_i}&=\Big|\frac{m_{l_i}}{f_a}\big[(k_f)_{ii}-(k_F)_{ii}\big]\Big|,\\
	c_{l_il_j}&=\sqrt{|(k_f)_{ij}|^2+|(k_F)_{ij}|^2},\qquad i\neq j.
\end{align}
The constraints on the diagonal couplings $g_{al_i}$ are dominated by stellar cooling processes~\cite{DILUZIO20201,Giannotti2017,Bollig2020}, 
while the bounds on the off-diagonal couplings $c_{l_i l_j}$ are mainly derived from the charged lepton flavor-violating (LFV) decay $l_i \to l_j a$~\cite{DILUZIO20201,Bauer_2022}. The most stringent limits currently read
\begin{equation}
	g_{ae} \lesssim 1.6\times10^{-13}, 
	\qquad 
	\frac{c_{\mu e}}{f_a}\lesssim 6.25\times10^{-10}\;\mathrm{GeV}^{-1}.
\end{equation}
Once the electron mass is factored out from $g_{ae}$, the two constraints become comparable in magnitude. Therefore, in the following estimate we use either constraint as a representative order-of-magnitude bound.

If the deviation of the PQ charge matrix from the identity
\begin{equation}
	\Delta Y=Y-1,
\end{equation}
is of $\mathcal{O}(1)$, it is reasonable to expect that the effective couplings differ from those of the flavor-universal models only by factors of order unity.

Consequently, under the assumption $\Delta Y=\mathcal O(1)$, the resulting bound is expected to be of the same order as that obtained in flavor-universal models,
\begin{equation}
	\frac{(y^{(D)})^2}{\lambda}
	\lesssim 10^{-8}\;\mathrm{GeV}^{-1},\label{eq:constraint}
\end{equation}
Assuming $\lambda\sim10^9\,$GeV, this constraint reduces to
\begin{equation}
	(y^{(D)})^2\lesssim10.
\end{equation}
This condition is comfortably satisfied for perturbative Yukawa couplings, $y^{(D)}\lesssim1$~\cite{Giannotti2017}. Therefore, the present model is naturally compatible with current experimental constraints on axion--charged lepton interactions.

\section{Conclusions}
\label{sec:conc}
In this work, we constructed a minimal flavored $\nu$KSVZ model by identifying the PQ symmetry with a lepton flavor symmetry. This identification strongly constrains the structure of the lepton sector. Imposing minimality in terms of the PQ charge assignment and the scalar content naturally leads to predictive patterns for the lepton mass matrices.

A particularly important consequence is that the inverse light-neutrino mass matrix $M_\nu^{-1}$ exhibits a characteristic zero texture. Confronting this texture with current neutrino oscillation data, we find that the model uniquely predicts an inverted neutrino mass ordering with $ 17\lesssim m_1/m_3\lesssim45 $.
Furthermore, the smallness of the mixing angle $\theta_{13}$ is naturally related to the hierarchy between the vacuum expectation values of the two PQ-charged scalar fields, providing a dynamical origin for this hierarchy. The effective Majorana mass $\langle m_{\beta\beta}\rangle$ is also predicted to satisfy a simple approximate relation proportional to $m_3^{-1}$.

The resulting neutrino phenomenology is highly predictive. In particular, the model predicts $\Sigma \simeq0.10~\mathrm{eV}$ and 	$\langle m_{\beta\beta}\rangle\simeq19\text{--}50~\mathrm{meV}$.
Current oscillation data show only a mild preference for the normal ordering, while cosmological constraints on $\Sigma$ remain dependent on the underlying cosmological model, and the present KamLAND-Zen bound excludes only part of the predicted $\langle m_{\beta\beta}\rangle$ interval. Consequently, although the model is already subject to nontrivial experimental tests, it is not excluded at present. Future neutrino oscillation, cosmological, and $0\nu\beta\beta$ experiments will provide decisive tests of this framework.

We also performed a systematic analysis of the axion--lepton interactions. We first identified the axion field and its decay constant in the present framework and then derived the leading-order effective interactions between the axion and leptons. In the neutrino sector, we discussed the constraints on the axion--neutrino couplings together with the possible modifications to neutrino oscillations, and presented a possible enhancement mechanism in \ref{ap:clock}.

In the charged-lepton sector, to make the one-loop calculation as model-independent as possible, we assumed only that the PQ charge matrix is diagonal and commutes with the Dirac neutrino mass matrix and the left-handed charged-lepton mixing matrix. Under these assumptions, we derived analytical expressions for the one-loop axion--charged lepton couplings, demonstrated explicitly the cancellation of the UV divergences, and showed that one class of diagrams vanishes irrespective of the scalar mixing pattern. Although the flavored framework predicts additional flavor--changing axion couplings compared with flavor-universal models, the diagonal CP-violating axion--charged lepton couplings remain zero at leading order, where contributions of $\mathcal{O}(\epsilon^2)$ are neglected. The resulting effective couplings are compatible with current experimental constraints.

Taken together, these results demonstrate that identifying the PQ symmetry with lepton flavor not only provides a viable solution to the strong CP problem, but also yields a predictive and experimentally testable framework connecting neutrino masses, flavor structure, and axion phenomenology.

\section*{Declaration of generative AI and AI-assisted technologies in the manuscript preparation process}
During the preparation of this work, the authors used ChatGPT in order to improve language and readability of the manuscript. After using this tool, the authors reviewed and edited the content as needed and take full responsibility for the content of the published article.

\section*{acknowledgments}
This work was supported in part by the Young Scientists Fund of the National Natural Science Foundation of China (Grant No. 11505067).

\appendix
	
\section{Proof about the PQ charge assignment}
\label{ap:proof}
In this appendix, we give a short derivation showing why the case $N_{PQ}=2$ cannot be realized and why, up to permutations of the generation indices, there is a unique nontrivial PQ charge assignment for the case $N_{PQ}=3$.

The PQ charges appearing in the Yukawa coefficients $x_{ijk}$, or equivalently in the bilinears $\overline{N^c_{iR}} N_{jR}$, are
\begin{equation}
	2\chi_1,\quad 2\chi_2,\quad 2\chi_3,\quad
	\chi_1+\chi_2,\quad \chi_1+\chi_3,\quad \chi_2+\chi_3 .
\end{equation}
The number $N_{PQ}$ denotes the number of distinct values among these six charges.

\begin{itemize}
\item $N_{PQ}=2$:

For $N_{PQ}=2$, at least two of the three charges $(\chi_1,\chi_2,\chi_3)$ must be different; otherwise the model reduces to the flavor-universal case $N_{PQ}=1$. Without loss of generality, let us assume
\begin{equation}
	\chi_1\neq \chi_2 .
\end{equation}
Then the three charge combinations
\begin{equation}
	2\chi_1,\quad 2\chi_2,\quad \chi_1+\chi_2
\end{equation}
are already distinct. Hence at least three distinct PQ charges necessarily appear in the Yukawa coefficients, and the case $N_{PQ}=2$ is algebraically impossible.
	
\item $N_{PQ}=3$:

For three generic charges $\chi_i$, the six combinations listed above are all distinct. To reduce the number of distinct charges to $N_{PQ}=3$, some of these combinations must be identified. The possible elementary identifications are of the form
\begin{equation}
	\chi_i=\chi_j,
	\quad
	2\chi_i=\chi_j+\chi_k,
\end{equation}
where $(i,j,k)$ denote three mutually different generation indices.

If two or more independent constraints of this type are imposed, the charge assignment collapses to the flavor-universal case. For example, the two constraints
\begin{equation}
	\chi_1+\chi_2=2\chi_3,
	\quad
	\chi_1+\chi_3=2\chi_2
\end{equation}
imply
\begin{equation}
	\chi_1=\chi_2=\chi_3 .
\end{equation}
The other choices can be analyzed in the same way and lead to the same flavor-universal result. Therefore, in a nontrivial flavored assignment with $N_{PQ}=3$, only one independent identification can be imposed.

If the single identification is of the type
\begin{equation}
	2\chi_i=\chi_j+\chi_k ,
\end{equation}
with $i,j,k$ mutually different, then the number of distinct charges is still larger than three. For instance, taking
\begin{equation}
	2\chi_3=\chi_1+\chi_2
\end{equation}
gives $\chi_3=(\chi_1+\chi_2)/2$, and the six charge combinations reduce to
\begin{equation}
	2\chi_1,\quad
	2\chi_2,\quad
	\chi_1+\chi_2,\quad
	\frac{3\chi_1+\chi_2}{2},\quad
	\frac{\chi_1+3\chi_2}{2}.
\end{equation}
For $\chi_1\neq\chi_2$, these are five distinct values. Thus this type of constraint cannot realize $N_{PQ}=3$.
	
The only remaining possibility is therefore a single equality between two of the three charges. Up to a permutation of the generation indices, we may take
\begin{equation}
	\chi_1\neq\chi_2=\chi_3.
\end{equation}
Then the six charge combinations reduce to three distinct values,
\begin{equation}
	2\chi_1,\quad
	2\chi_2,\quad
	\chi_1+\chi_2 .
\end{equation}
Equivalently, the charge matrix of the bilinears takes the form
\begin{equation}
	\chi(\overline{N^c_{iR}} N_{jR})=\begin{pmatrix}
		2\chi_1&\chi_1+\chi_2&\chi_1+\chi_2\\
		\chi_1+\chi_2&2\chi_2&2\chi_2\\
		\chi_1+\chi_2&2\chi_2&2\chi_2
	\end{pmatrix},
\end{equation}
All other assignments of this type are related to this one by permutations of the generation labels. Therefore, up to such permutations, there is a unique nontrivial PQ charge assignment in the $N_{PQ}=3$ scenario.
\end{itemize}

\section{Numerical verification of the analytical results}
\label{ap:num}

In this appendix, we provide a numerical verification to complement the analytical discussion in Sec.~\ref{sec:hierarchy}. We adopt the central values and the uncertainties from the current neutrino oscillation data \cite{PDG2024}:
\begin{align}
	\Delta m_{21}^2 &= (7.53\pm0.18)\times10^{-5}~\text{eV}^2, 
	& \sin^2\theta_{12} &= 0.307\pm0.013, \nonumber\\
	\Delta m_{32}^2 &= (-2.529\pm0.029)\times 10^{-3}~\text{eV}^2, 
	& \sin^2\theta_{23} &= 0.553^{+0.016}_{-0.024}, \nonumber\\
	\delta_{13}&=1.19\pm 0.22~ \pi~\text{rad} & \sin^2\theta_{13} &= (2.19\pm0.07)\times 10^{-2}.
	\label{eq:PMNSdata}
\end{align}
Notably, Eq.~\eqref{eq:texture3} depends on only two independent phase combinations, $\eta_1+\delta_{13}$ and $\eta_2+\delta_{13}$. In our numerical procedure, we vary these two phases within the interval $[0, \pi]$. By taking $\eta_2+\delta_{13}$ as an input parameter, the values of $m_3$, $\eta_1+\delta_{13}$, and the effective mass $\langle m_{\beta\beta}\rangle$ can be determined numerically by solving the texture condition in Eq.~\eqref{eq:texture3}.

\begin{figure}
\centering
\begin{minipage}[t]{0.48\textwidth}
	\includegraphics[width=\linewidth]{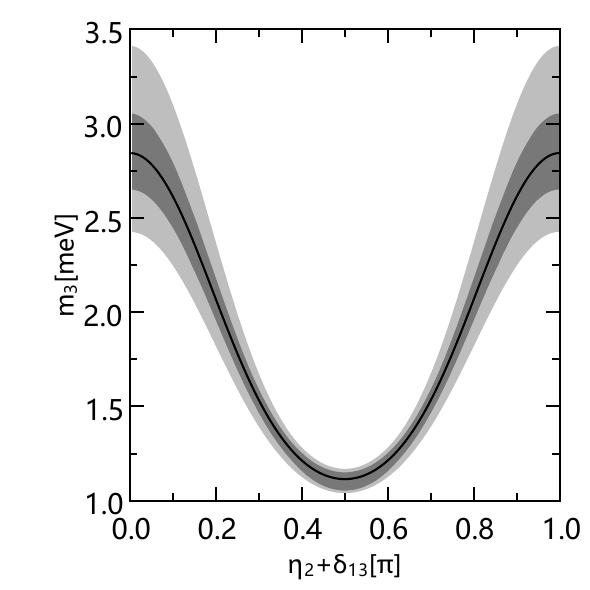}
\end{minipage}
\hfill
\begin{minipage}[t]{0.48\textwidth}
	\includegraphics[width=\linewidth]{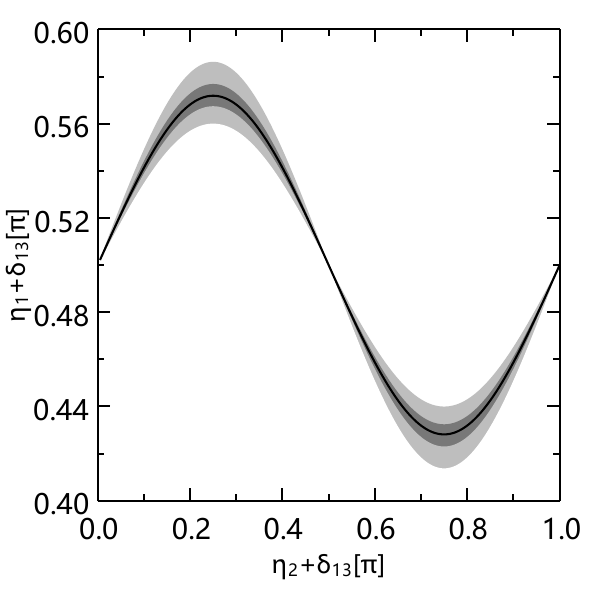}
\end{minipage}
\caption{The lightest neutrino mass $m_3$ (left panel) and the phase combination $\eta_1+\delta_{13}$ (right panel) plotted against $\eta_2+\delta_{13}$. The solid black curves represent the results using the central values of the oscillation parameters, while the dark (light) gray regions depict the $1\sigma$ ($3\sigma$) experimentally allowed ranges based on Ref.~\cite{PDG2024}.}
	\label{fig:m3andeta1}
\end{figure}

In the left panel of Fig.~\ref{fig:m3andeta1}, we present $m_3$ as a function of the phase combination $\eta_2+\delta_{13}$, while the right panel displays the correlation between $\eta_1+\delta_{13}$ and $\eta_2+\delta_{13}$. The solid black curves are obtained by fixing all mixing parameters at their best-fit values, while the gray regions account for their experimental uncertainties

It is worth highlighting that two special points emerge in the right panel of Fig.~\ref{fig:m3andeta1}:
\begin{align*}
	&1)\eta_2+\delta_{13}=\eta_1+\delta_{13}=\frac{\pi}{2},\\
	&2)\eta_1+\delta_{13}=\frac{\pi}{2},\; \eta_2+\delta_{13}=0.
\end{align*}
One observes that as the phase configuration approaches these two points, the area of the gray region significantly shrinks. This phenomenon is consistent with our analytical results in Sec.~\ref{sec:IH}, confirming that these two points represent exact solutions that are independent of other mixing parameters.

This property also elucidates why the numerical results in Ref.~\cite{Lashin2009} appear to exclude the region near $\eta_1 = \eta_2$. In a standard numerical scan, the probability of random sampling points falling within a given region is proportional to its allowed parameter space area. Near $\eta_1 = \eta_2$, the ``width'' of the gray region (induced by experimental uncertainties) approaches zero, making it much less likely to be populated by random points compared to other regions. Consequently, such points may be mistakenly overlooked in a purely numerical treatment.

\begin{figure}
\centering
\begin{minipage}[t]{0.48\textwidth}
	\includegraphics[width=\linewidth]{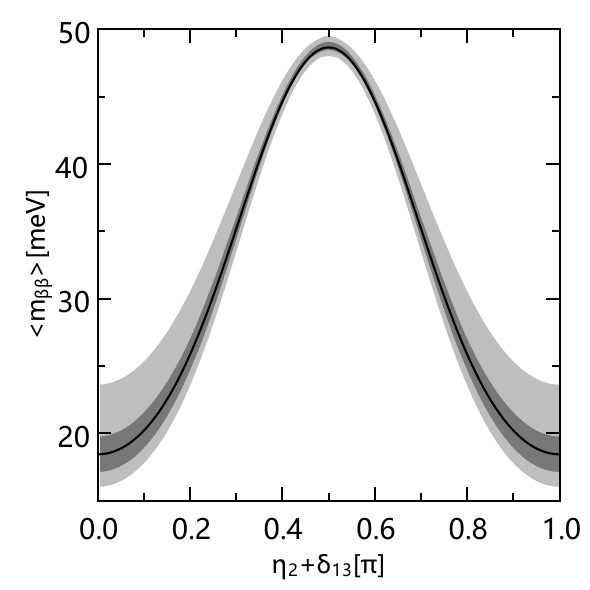}
\end{minipage}
\hfill
\begin{minipage}[t]{0.48\textwidth}
	\includegraphics[width=\linewidth]{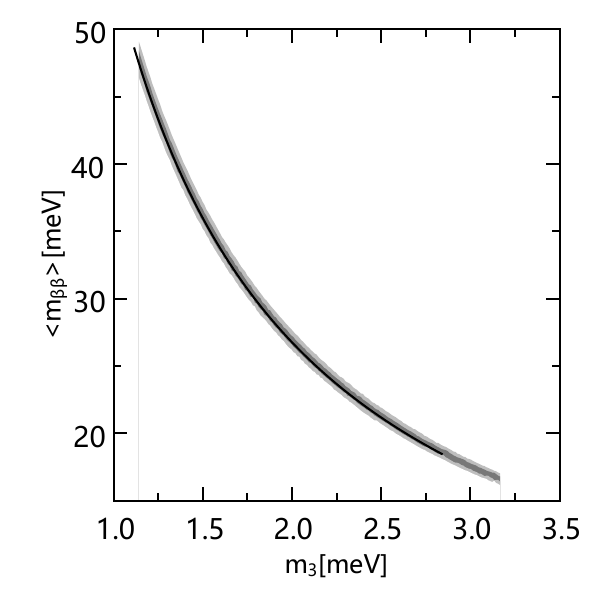}
\end{minipage}
\caption{The effective Majorana mass $\langle m_{\beta\beta}\rangle$ plotted against $\eta_2+\delta_{13}$ (left panel) and $m_3$ (right panel). The solid black curves represent the results using the central values of the oscillation parameters, while the dark (light) gray regions depict the $1\sigma$ ($3\sigma$) experimentally allowed ranges based on Ref.~\cite{PDG2024}.}
\label{fig:mee}
\end{figure}

In Fig. \ref{fig:mee}, we present the correlations of $\langle m_{\beta\beta}\rangle$ as a function of $\eta_2+\delta_{13}$ (left panel) and its dependence on the lightest neutrino mass $m_3$ (right panel). Both numerical results show excellent agreement with the analytical derivations provided in Sec.~\ref{sec:IH}.

To illustrate the hierarchy among the matrix elements of $M^{(R)}$, we assume the following values for simplicity
\begin{equation}
	U_{lL}=1,\quad M^{(D)}=1~\mathrm{GeV}.
\end{equation}
Note that this choice of $M^{(D)}$ is intended only to illustrate the relative structure of $M^{(R)}$ and does not serve as a physical benchmark for the seesaw scale.
By substituting the numerical solutions for $(\eta_2, m_3, \eta_1)$ obtained from the texture condition into Eq.~\eqref{eq:hierM}, we can determine the specific magnitudes of the elements in $M^{(R)}$.

\begin{figure}[htbp]
\centering

\begin{minipage}[t]{0.48\textwidth}
	\centering

	\includegraphics[height=0.21\textheight, width=\linewidth, keepaspectratio]{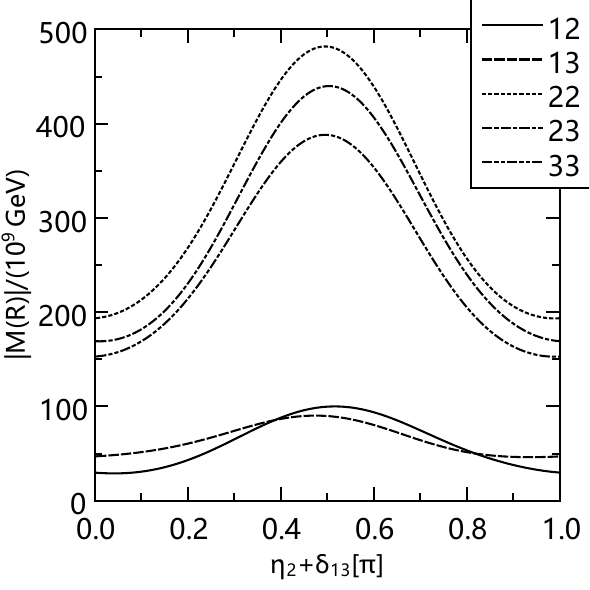}
	\centerline{\footnotesize (a)}
\end{minipage}
\hfill
\begin{minipage}[t]{0.48\textwidth}
	\centering
	\includegraphics[height=0.21\textheight, width=\linewidth, keepaspectratio]{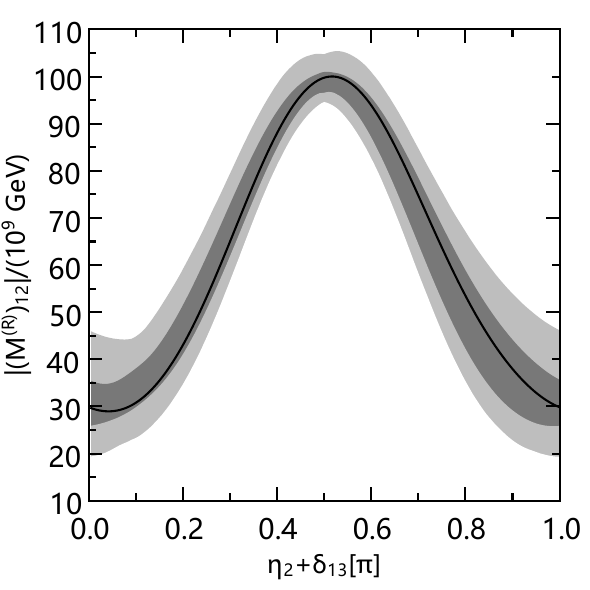}
	\centerline{\footnotesize (b)}
\end{minipage}

\vspace{0.2cm} 
	
\begin{minipage}[t]{0.48\textwidth}
	\centering
	\includegraphics[height=0.21\textheight, width=\linewidth, keepaspectratio]{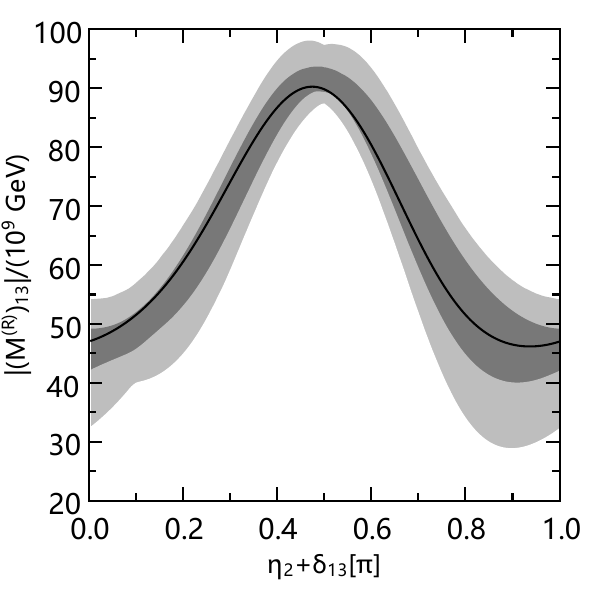}
	\centerline{\footnotesize (c)}
\end{minipage}
\hfill
\begin{minipage}[t]{0.48\textwidth}
	\centering
	\includegraphics[height=0.21\textheight, width=\linewidth, keepaspectratio]{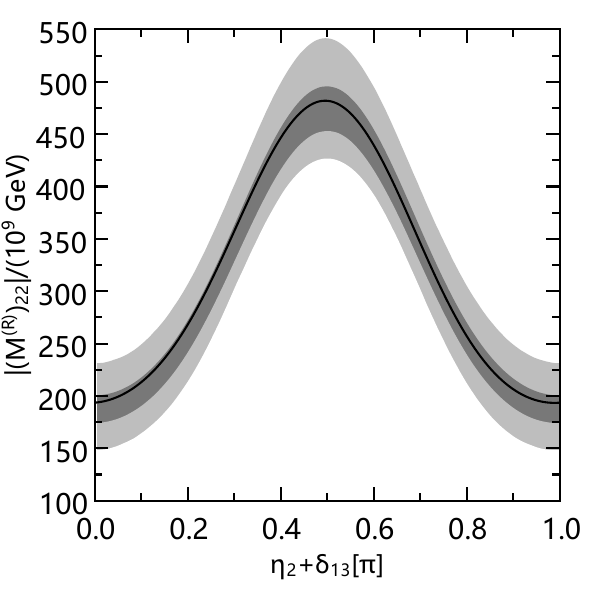}
	\centerline{\footnotesize (d)}
\end{minipage}
	
\vspace{0.2cm}
	
\begin{minipage}[t]{0.48\textwidth}
	\centering
	\includegraphics[height=0.21\textheight, width=\linewidth, keepaspectratio]{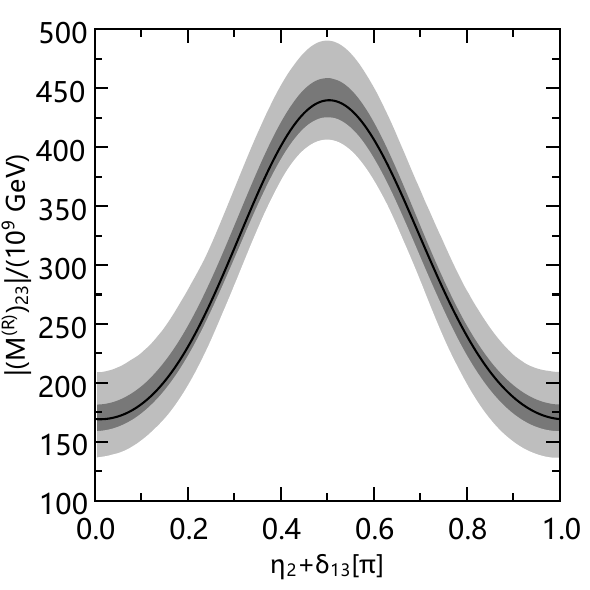}
	\centerline{\footnotesize (e)}
\end{minipage}
\hfill
\begin{minipage}[t]{0.48\textwidth}
	\centering
	\includegraphics[height=0.21\textheight, width=\linewidth, keepaspectratio]{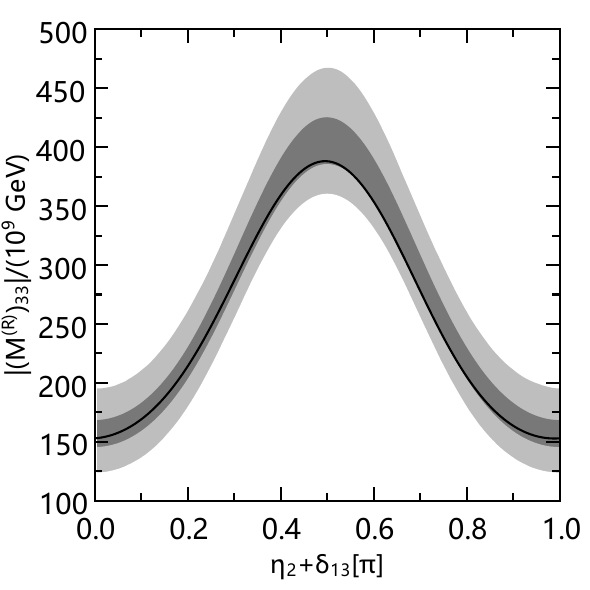}
	\centerline{\footnotesize (f)}
\end{minipage}
	
\caption{Numerical results for the matrix elements of $M^{(R)}$ as functions of $\eta_2+\delta_{13}$. Panel (a) compares the absolute values of all five elements using central values. In panels (b)--(f), the solid black curves represent the results for the central values of mixing parameters, while the dark (light) gray regions depict the $1\sigma$ ($3\sigma$) experimental uncertainties from Ref.~\cite{PDG2024}.}
\label{fig:MR_all_6}
\end{figure}

Finally, the numerical distributions of the $M^{(R)}$ matrix elements are summarized in Fig.~\ref{fig:MR_all_6}. Panel (a) compares the magnitudes of all five non-zero elements based on the central values of the mixing parameters. To illustrate the impact of experimental precision, panels (b) through (f) present the individual correlations for each element against the phase $\eta_2+\delta_{13}$.

\section{Clockwork enhancement}
\label{ap:clock}

In this appendix, we describe how the clockwork mechanism \cite{PhysRevD.103.015034} can be used to generate a large PQ charge difference $\chi-1$.

To implement this mechanism, we introduce a chain of SM-singlet scalar fields
$S_1, S_2, \dots, S_n$, with the last field identified as
\begin{equation}
	S_n=\phi_1 .
\end{equation}
Their PQ charges are denoted by $-\chi_{S_1},-\chi_{S_2},\dots,-\chi_{S_n}=-\chi-1$.  
All fields $S_i$ except $S_n$ do not couple directly to leptons or heavy neutrinos, and therefore the lepton mass matrices derived in the main text remain unchanged.

Since there are $n+1$ singlet scalars $S_1,\dots,S_n,\phi_2$, one needs $n$ explicit symmetry-breaking terms to reduce
\begin{equation}
	U(1)^{n+1}\rightarrow U(1)_{\rm PQ}.
\end{equation}
We choose $n-1$ of them to be
\begin{equation}
	S_{k-1}^3 S_k^{*},\qquad k=2,3,\dots,n.
\end{equation}
These terms fix the relations among the PQ charges of neighboring fields. 
Iterating these relations yields the charge relation between $\phi_1$ and $S_1$,
\begin{equation}
	1+\chi=3^{\,n-1}\chi_{S_1} .
\end{equation}

The final breaking term can be chosen as
\begin{equation}
	\phi_1^2 \phi_2^{*} S_1^{*},
\end{equation}
which implies $\chi_{S_1}=2$. 
Consequently, the PQ charge $\chi$ is determined by the number of scalar fields $n$,
\begin{equation}
	\chi=-1+2\cdot 3^{\,n-1}\sim 3^{\,n}.
	\label{eq:chi}
\end{equation}

As shown in Sec.~\ref{sec:AxionNu}, observable modifications of axion--neutrino interaction require
\begin{equation}
	\chi \sim 10^9 .
\end{equation}
This estimate indicates that a clockwork chain with roughly
\begin{equation}
	n \sim 19
\end{equation}
scalar fields is sufficient to enhance the axion--neutrino coupling to a level potentially accessible to future experiments.

On the other hand, solving the strong CP problem requires a nonvanishing QCD anomaly coefficient.  
With the PQ charge assignments introduced above, this condition can be readily satisfied. 
For instance, introducing a Yukawa interaction between a heavy quark $Q$ and the scalar $S_1$,
\begin{equation}
	\bar{Q}_L Q_R S_1 + \text{h.c.},
\end{equation}
after the field-dependent transformation $Q\rightarrow \mathrm{e}^{\mathrm{i}\gamma_5a/\lambda}Q$, generates a color anomaly term
\begin{equation}
	-\frac{2a}{\lambda}\frac{g_s^2}{32\pi^2}G\tilde{G}.
\end{equation}
However, the same clockwork enhancement that amplifies the axion--neutrino coupling also enhances the axion--charged lepton couplings.  
For $\chi \sim 10^9$, the constraint \eqref{eq:constraint} implies
\begin{equation}
	y^{(D)} \lesssim 10^{-4}.
\end{equation}
Via the seesaw relation~\eqref{eq:seesaw1}, this upper bound on $y^{(D)}$ translates into an upper bound on the heavy neutrino masses,
\begin{equation}
	m_N \lesssim 10^6~\text{GeV}.
\end{equation}
Since heavy neutrinos with masses of order $10^6~\text{GeV}$ remain well within current experimental and cosmological bounds, the clockwork enhancement considered here is therefore phenomenologically viable.

\section{The loop integrals}
\label{ap:calc}
We denote $x$ and $y$ as the Feynman parameters, then
\begin{align}
	I_1(\lambda_l,\lambda_k)=&\int_0^1\mathrm{d}x\int_0^{1-x}\mathrm{d}y \; \frac{1}{x(\lambda_l-1)+y(\lambda_k-1)+1}\notag\\
	=&\frac{\lambda_k\ln\lambda_k}{(\lambda_k-1)(\lambda_k-\lambda_l)}+\frac{\lambda_l\ln\lambda_l}{(\lambda_l-1)(\lambda_l-\lambda_k)},\\
	I_2(\lambda_l,\lambda_k)=&\int_0^1\mathrm{d}x\int_0^{1-x}\mathrm{d}y \; \frac{x}{x(\lambda_l-1)+y(\lambda_k-1)+1},\notag\\
	=&\frac{\lambda_k^2\ln\lambda_k}{2(\lambda_k-1)(\lambda_k-\lambda_l)^2}-\frac{\lambda_l\big[\lambda_l+\lambda_k(\lambda_l-2)\big]\ln\lambda_l}{2(\lambda_l-1)^2(\lambda_k-\lambda_l)^2}
	-\frac{\lambda_l}{2(\lambda_k-\lambda_l)(\lambda_l-1)},\\
	I_3(\lambda_l,\lambda_k)=&\int_0^1\mathrm{d}x\int_0^{1-x}\mathrm{d}y \; \frac{y}{x(\lambda_l-1)+y(\lambda_k-1)+1}\notag\\
	=&\frac{1}{2(\lambda_k-1)}-\frac{\lambda_l-1}{\lambda_k-1}I_2(\lambda_l,\lambda_k)-\frac{1}{\lambda_k-1}I_1(\lambda_l,\lambda_k),\\
	C_{\text{fin}}(\lambda_l,\lambda_k)=&\int_0^1\mathrm{d}x\int_0^{1-x}\mathrm{d}y \; \ln[x(\lambda_l-1)+y(\lambda_k-1)+1]\notag\\
	=&\frac{2\lambda_l^2(\lambda_k-1)\ln\lambda_l-2\lambda_k^2(\lambda_l-1)\ln\lambda_k-3(\lambda_l-1)(\lambda_k-1)(\lambda_l-\lambda_k)}{4(\lambda_l-1)(\lambda_k-1)(\lambda_l-\lambda_k)},\\
	L_0(\lambda_l,\lambda_k)=&\int_0^1\mathrm{d}x \; \ln[x\lambda_l+(1-x)\lambda_k]=\frac{\lambda_l\ln\lambda_l}{\lambda_l-\lambda_k}+\frac{\lambda_k\ln\lambda_k}{\lambda_k-\lambda_l}-1,\\	
	L_1(\lambda_l,\lambda_k)=&\int_0^1\mathrm{d}x \; x\ln[x\lambda_l+(1-x)\lambda_k]
	=-\frac{1}{4}+\frac{\lambda_k}{2(\lambda_l-\lambda_k)}+\frac{\lambda_k^2}{2(\lambda_l-\lambda_k)^2}\ln\Big(\frac{\lambda_k}{\lambda_l}\Big)+\frac{1}{2}\ln\lambda_l.
\end{align}

\section{Proof of the matrix identities}
\label{ap:id}
In this appendix, we prove the matrix identities in Eqs.~\eqref{eq:id1}--\eqref{eq:id4}. We first rewrite $B$, $C$, and $\mathcal{Y}$ in matrix-product form using their definitions:
\begin{equation}
	B=U_{lL}^\dagger\begin{pmatrix}
		1&0
	\end{pmatrix}U_n^*,\quad C=U_n^T\begin{pmatrix}
	1&0\\
	0&0
	\end{pmatrix}U_n^*,\quad \mathcal{Y}=U_n^\dagger\begin{pmatrix}
	-Y&0\\
	0&Y
	\end{pmatrix}U_n,
\end{equation}
where each block of the middle matrices is a $3\times3$ matrix.

Using these expressions, the identities in Eqs.~\eqref{eq:id1}--\eqref{eq:id4} can be proved straightforwardly:
\begin{itemize}
	\item[1.] $B\mathcal{Y}^*=-YB$
	\begin{equation}
		B\mathcal{Y}^*=-U_{lL}^\dagger\begin{pmatrix}
			Y&0
		\end{pmatrix}U_n^*=-YB,
	\end{equation}
	where we use the assumption \eqref{eq:assumptionb} in the second equality.
	\item[2.] $B\hat{M}_n\mathcal{Y}=YB\hat{M}_n$.\\
	\begin{align}
		B\hat{M}_n\mathcal{Y}&=U_{lL}^\dagger\begin{pmatrix}
			1&0
		\end{pmatrix}M_n\begin{pmatrix}
		-Y&0\\
		0&Y
		\end{pmatrix}U_n
		=YU_{lL}^\dagger\begin{pmatrix}
		1&0
		\end{pmatrix}M_nU_n=YB\hat{M}_n,
	\end{align}
	where the first and last equalities follow from Eq.~\eqref{eq:UnMn}, while the second equality uses the assumption \eqref{eq:assumptionb}.
	\item[3.] $C\mathcal{Y}^*=\mathcal{Y}^*C$
	\begin{equation}
			\mathcal{Y}^*C=-U_n^T\begin{pmatrix}
				Y&0\\
				0&0
			\end{pmatrix}U_n^*=C\mathcal{Y}^*.
	\end{equation}
	\item[4.] $C\hat{M}_n\mathcal{Y}=-C\mathcal{Y}^*\hat{M}_n$
	\begin{align}
		C\hat{M}_n\mathcal{Y}&=U_n^T\begin{pmatrix}
			1&0\\
			0&0
		\end{pmatrix}M_n\begin{pmatrix}
			-Y&0\\
			0&Y
		\end{pmatrix}U_n
		=-U_n^T\begin{pmatrix}
			1&0\\
			0&0
		\end{pmatrix}\begin{pmatrix}
			-Y&0\\
			0&Y
		\end{pmatrix}M_nU_n=-C\mathcal{Y}^*\hat{M}_n,
	\end{align}
	where the first and last equalities follow from Eq.~\eqref{eq:UnMn}, while the second equality uses the assumption \eqref{eq:assumptionb}.
\end{itemize}
Finally, we compare these identities with those derived in Ref.~\cite{Pilaftsis_1994} (also discussed in Ref.~\cite{Heeck2019}), since both sets of identities play a crucial role in establishing the cancellation of the ultraviolet divergences.

It can be verified that by taking $Y\propto 1$, the identities \eqref{eq:id1}, \eqref{eq:id2} and \eqref{eq:id4} reduce to
\begin{align}
	B\mathcal{Y}^*=-YB \;&\Rightarrow\; BC=B,\\
	B\hat{M}_n\mathcal{Y}=YB\hat{M}_n \;&\Rightarrow\; B\hat{M}_nC^*=0,\\
	C\hat{M}_n\mathcal{Y}=-C\mathcal{Y}^*\hat{M}_n \;&\Rightarrow\; C\hat{M}_nC^*=0.
\end{align}
These identities coincide exactly with Eq.~(27), Eq.~(29) and Eq.~(30) of Ref.\cite{Pilaftsis_1994}.
The reduction of Eq.~\eqref{eq:id3} is immediate, since it simply states that $C$ commutes with itself. The identity corresponding to Eq.~(28) of Ref.~\cite{Pilaftsis_1994} also holds in the present framework:
\begin{equation}
	C^2=U_n^T\begin{pmatrix}
		1&0\\
		0&0
	\end{pmatrix}\begin{pmatrix}
	1&0\\
	0&0
\end{pmatrix}U_n^*=C.
\end{equation}
Therefore, the identities in Eqs.~\eqref{eq:id1}--\eqref{eq:id4} provide the natural generalization of the flavor-universal identities derived in Ref.~\cite{Pilaftsis_1994}. As in the flavor-universal case, they play a central role in establishing the ultraviolet finiteness of amplitudes involving neutrino loops~\cite{Heeck2019}.

\bibliographystyle{elsarticle-num} 
\bibliography{axion_neutrino.bib}

\end{document}